\documentclass[aps,twocolumn,groupedaddress,showpacs,prb]{revtex4}
\usepackage{graphicx,rotating,subfigure,amsmath,amsfonts,amssymb,delarray}
\renewcommand{\vec}[1]{\boldsymbol #1}
\newcommand{\e}{\text{e}}
\newcommand{\im}{\text{i}}

\begin{document}
\bibliographystyle{apsrev}


\title{Thermodynamics and Crossover Phenomena in the Correlation Lengths of the One-Dimensional t-J Model}


\author{J. Sirker}
\email[]{sirker@fkt.physik.uni-dortmund.de}
\affiliation{Theoretische Physik I, Universit\"at Dortmund, Otto-Hahn-Str.\!\! 4, D-44221 Dortmund, Germany}

\author{A. Kl\"umper}
\email[]{kluemper@fkt.physik.uni-dortmund.de}
\affiliation{Theoretische Physik I, Universit\"at Dortmund, Otto-Hahn-Str.\!\! 4, D-44221 Dortmund, Germany}


\date{\today}

\begin{abstract}
We investigate the thermodynamics of the one-dimensional t-J model using transfer matrix 
renormalization group (TMRG) algorithms and present results for quantities
like particle number, specific heat, spin susceptibility and
compressibility. Based on these results we confirm a phase diagram 
consisting of a Tomonaga-Luttinger liquid (TLL) phase for small J/t and a
phase separated state for J/t large. Close to phase separation
we find a spin-gap (Luther-Emery) phase at low densities consistent with
predictions by other studies. At the supersymmetric point we
compare our results with exact results from the Bethe ansatz and find excellent
agreement. In particular we focus on the calculation of correlation lengths
and static correlation functions and study the crossover from the
non-universal high T lattice into the quantum critical regime.  At the
supersymmetric point we compare in detail with predictions by conformal field theory (CFT) and TLL theory and show the importance of logarithmic corrections. 
\end{abstract}

\maketitle

\section{Introduction}
\label{Intro}
The t-J model is one of the most fundamental systems of strongly correlated electrons. The two-dimensional version has attracted much attention because it is believed that it describes the basic interactions in the copper-oxygen planes of high-$T_c$ superconductors. 
For the one-dimensional (1D) t-J model much progress has been achieved using various analytical and numerical techniques.\cite{OgataLuchini,HellbergMele,BaresBlatter,KawakamiYang,JuettnerKluemper,ChenLee,KobayashiOhe,NakamuraNomura} At the supersymmetric point $J/t=2$ the model is solvable by the Bethe ansatz and ground state properties as well as the excitation spectra have been obtained exactly.\cite{BaresBlatter} Because the two critical excitations of spin and charge type are separated, the properties can be described by two independent $c=1$ Virasoro algebras. By a combination of finite-size results from the Bethe ansatz and conformal field theory (CFT) it is therefore also possible to calculate the critical exponents of algebraically decaying correlation functions.\cite{KawakamiYang} This explicitly shows that the t-J model at the supersymmetric point behaves as a Tomonaga-Luttinger liquid (TLL) for all electron densities. Thermodynamic quantities at this special point have been obtained by thermodynamic Bethe ansatz \cite{OkijiSuga} as well as by a combination of a Trotter-Suzuki mapping leading to a quantum transfer matrix (QTM) and the Bethe ansatz.\cite{JuettnerKluemper} Exact results are also available in the limit $J/t\rightarrow 0$ where the t-J model is equivalent to the Hubbard model with $U/t\rightarrow \infty$, showing again TLL behavior.\cite{FrahmKorepin} It is therefore believed that the t-J model shows TLL properties for all $0\leq J/t \leq 2$, what is supported by various numerical calculations.\cite{OgataLuchini,HellbergMele,ChenLee,KobayashiOhe} In these numerical works there is also general agreement that the t-J model phase separates for $J/t=2.8$ to $3.5$ depending on the electron density. 

Already Ogata {\it et al.}~\cite{OgataLuchini} conjectured a third phase with
a spin gap in the low-density region for $J/t>2$. However, by calculating the
spin susceptibility for an electron density $n=1/3$ on small chains no
evidence for a spin gap was found. Also the variational quantum Monte Carlo
(QMC) calculations of Hellberg and Mele \cite{HellbergMele} could not confirm
the appearance of a spin gap. By using the same method but other trial wave
functions a phase with Luther-Emery (LE) properties was found by Chen and Lee
\cite{ChenLee} and Kobayashi {\it et al.}~\cite{KobayashiOhe} at low densities and $2<J/t<3.1$. However, the obtained result strongly depends on the trial wave functions used in the calculations. 
Completely different phase boundaries with the spin-gap phase extending into the high density region have been obtained by Nakamura {\it et al.}~\cite{NakamuraNomura} using a renormalization group treatment of the Tomonaga-Luttinger model under the assumption that the spin gap is caused by an attractive backward scattering process. They argue that the spin gap was underestimated in the numerical calculations, because it is the result of a marginal operator leading to an exponentially small gap. 

In this paper we want to examine thermodynamics of the one-dimensional t-J model in the whole $J/t$-parameter region. In particular, we are interested in the temperature dependence of correlation lengths. We use a grandcanonical description of the model, where the Hamiltonian is given by
\begin{eqnarray}
\label{t-J_Ham} 
H &=& -t\sum_{i,\sigma} {\it P} (c^{\dagger}_{i,\sigma} c_{i+1,\sigma} + c^{\dagger}_{i+1,\sigma} c_{i,\sigma}) {\it P} \\
&+& J\sum_i \left( \vec{S}_i \vec{S}_{i+1} - \frac{n_i n_{i+1}}{4}\right) - h \sum_i S^z_i - \mu \sum_i n_i \nonumber
\end{eqnarray}
with a magnetic field $h$, a chemical potential $\mu$ and ${\it P}$ being the projection operator onto the Hilbert-subspace without double occupancy. To study thermodynamic properties also away from the supersymmetric point the TMRG provides a powerful numerical tool. This method is particularly suited, because the thermodynamic limit is performed exactly and it has been applied successfully to various one-dimensional systems before.\cite{BursillXiang,WangXiang,Shibata,Raupach} Compared with QMC methods it has further the advantage of never suffering under the fermion sign problem restricting QMC often to relatively high temperatures.

In Sec.~\ref{TMRG} we give a brief introduction into the calculation of thermodynamic quantities by the TMRG algorithm and point to a novel algorithm with several advantages. Our results for the supersymmetric point are shown in Sec.~\ref{Susy} where we also compare the numerics with the Bethe ansatz results by J\"uttner {\it et al.}~\cite{JuettnerKluemper} In particular we calculate for this special point the temperature dependence of various correlation lengths and static correlation functions and discuss the low-temperature behavior in comparison to CFT predictions by Kawakami and Yang \cite{KawakamiYang} in detail. Because the t-J model is realized by the Hubbard model in the limit $U\gg t$ the case $J=2t^2/|U| < t$ is physically very relevant. Results for $J/t=0.35$, a value often used in literature, are given in Sec.~\ref{PhysRel}. As already mentioned the t-J model phase separates for $J/t$ large. The meaning of this in the grand-canonical ensemble is explained in Sec.~\ref{Phasesep}. Without using any assumption about the ground state or the low energy effective theory, the existence of a LE phase is proven in Sec.~\ref{Luther-Emery} by calculating directly spin susceptibilities as well as spin-spin and density-density correlation lengths. In Sec.~\ref{Aniso} we study the t-J model with an additional Ising-like anisotropy. The final section is devoted to our conclusions and in appendix \ref{Appendix} we compare numerical and exact results for the density-density correlation function of free spinless fermions.
\section{TMRG}
\label{TMRG}
After a decomposition of the Hamiltonian $H$ into even ($H_e$) and odd parts
($H_o$) the partition function is expressed by means of the Trotter formula
\begin{equation}
\label{Trotter-Suzuki}
Z = \text{Tr}\e^{-\beta H} = \lim_{M \rightarrow \infty}\text{Tr}\left\{\left[\e^{-\epsilon H_e}\e^{-\epsilon H_o}\right]^M\right\} 
\end{equation}
with $\epsilon =\beta/M$ and $\beta$ being the inverse temperature leading to a classical model on a lattice with checkerboard structure. The column-to-column transfer matrix (QTM),$T_M$, is a non-symmetric matrix describing the evolution along the spatial direction. The thermodynamic limit with fixed Trotter number M is performed exactly, because the free energy of the infinite chain is given solely by the largest eigenvalue $\Lambda_0$ of this QTM 
\begin{equation}
\label{free-energy}
f_{\infty,M} = -\frac{T}{2}\ln \Lambda_0 \; ,
\end{equation}
where $\Lambda_0$ is unique and a real, positive number for all temperatures. Since a vanishing gap between leading and next-leading eigenvalue indicates a phase transition, such a degeneracy is not possible for a 1D quantum system at finite temperature. The calculations are simplified by the conservation laws for spin and particle number or equivalently for the number of particles with spin up ($N_\uparrow$) and spin down ($N_\downarrow$) leading to a block structure of $T_M$. Expectation values of local operators can be expressed in terms of the left and right eigenvectors belonging to the largest eigenvalue and calculated directly. This is used for the inner energy $u$ and the magnetization $m$.
The behavior of a two-point correlation function is given by
\begin{equation}
\label{Corr-function}
\left< A_0 A_r \right> = \sum_\alpha M_{\alpha} \e^{-r/\xi_\alpha} \e^{\im k_\alpha r} 
\end{equation}
with $M_\alpha$ being matrixelements. The correlation lengths $\xi_\alpha$ can be calculated from the QTM by 
\begin{equation}
\label{corr-length}
\xi_\alpha^{-1} = \frac{1}{2} \ln \left| \frac{\Lambda_0}{\Lambda_\alpha} \right| 
\end{equation}
and the wavevectors $k_\alpha$ by
\begin{equation}
\label{wavevector}
k_\alpha = \frac{1}{2} \arg \left( \frac{\Lambda_\alpha}{\Lambda_0} \right) + n\pi \quad (n=0 \;\mbox{or}\; n=1), 
\end{equation}
where $\Lambda_\alpha$ are eigenvalues in the block of the QTM with non-vanishing matrixelements $M_\alpha$. This condition is controlled by a direct evaluation of
\begin{equation}
\label{matrixel}
M_\alpha = \frac{\langle \Psi_0^L | T_M(A_0) | \Psi_\alpha^R\rangle \langle \Psi_\alpha^L | T_M(A_r) | \Psi_0^R\rangle}{\Lambda_0\Lambda_\alpha},
\end{equation}
where $\langle \Psi_0^L |$, $| \Psi_0^R \rangle$ are the left and right eigenvectors belonging to the largest eigenvalue $\Lambda_0$ and $\langle \Psi_\alpha^L |$, $| \Psi_\alpha^R \rangle$ the left and right eigenvectors belonging to $\Lambda_\alpha$. 
The value for the wavevector $k_\alpha$ is not unique, because in the checkerboard decomposition of the partition function a local transfer matrix covers 2 sites. To overcome this ambiguity we have also applied a different Trotter-Suzuki mapping leading to a classical lattice with alternating rows. Because the transfer matrix in this formalism can be formulated for a single column, the wavevector in Eq.~(\ref{wavevector}) is determined unambiguously. Details of this modified TMRG algorithm will be published elsewhere.\cite{SirkerKluemper2} We have convinced ourselves that the accuracy of both algorithms is of the same order and use both of them in the following.
During all calculations we use the infinite DMRG algorithm to increase the Trotter number $M$ being equivalent to a decrease of the temperature $T$. The number of states kept in the DMRG varies between 64 and 240 and the fixed parameter $\epsilon$ between $0.025$ and $0.05$. 
\section{The supersymmetric point}
\label{Susy}
\begin{figure}[!ht]
\includegraphics*[width=0.9\columnwidth]{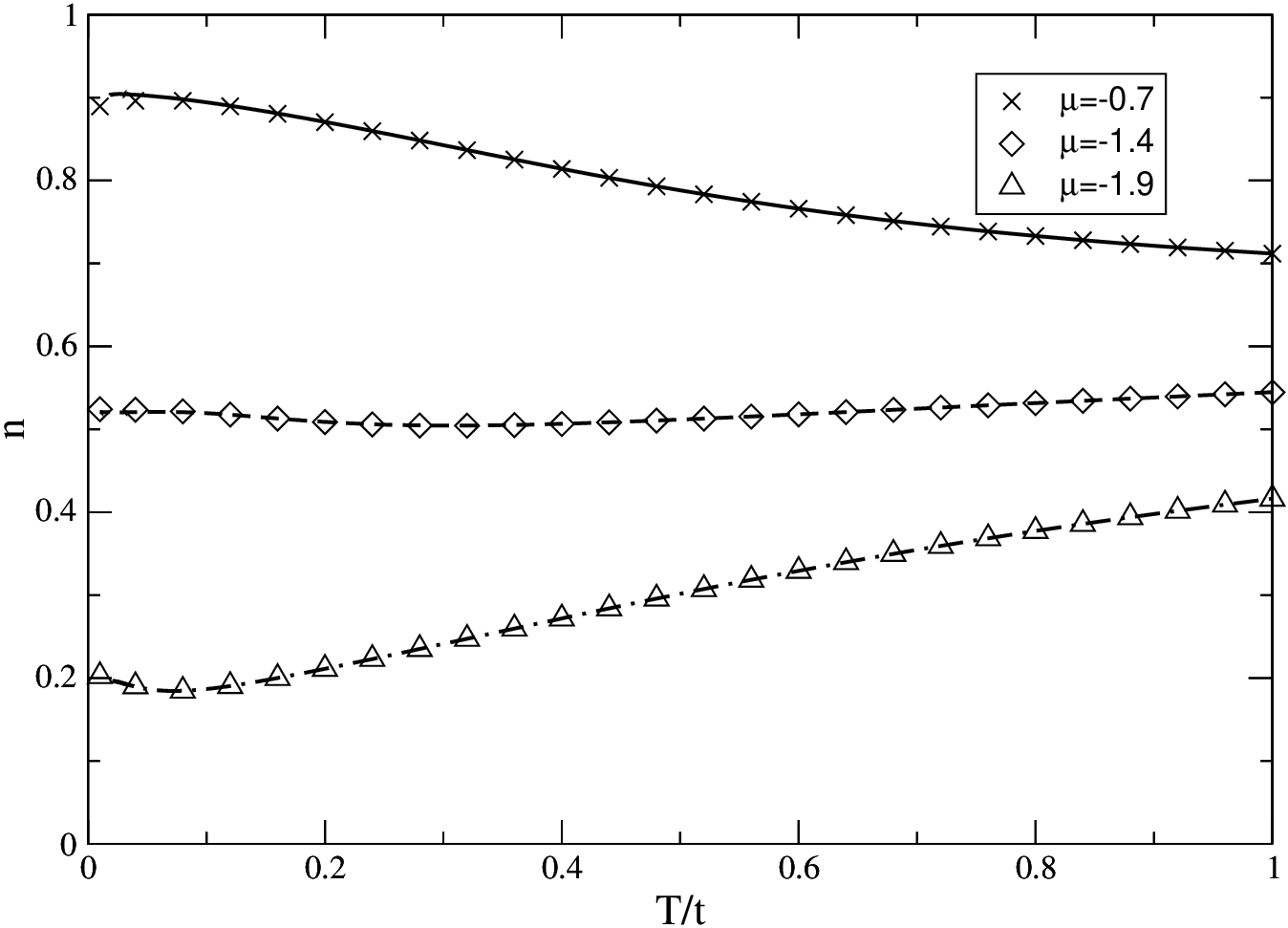}
\caption{Temperature dependence of the density for three different chemical potentials. The lines are given by the TMRG results, whereas the symbols denote the results from the Bethe ansatz.}
\label{fig_density}
\end{figure}

To check the numerics we have calculated several thermodynamic quantities for the supersymmetric point $J/t=2$ and compared with the exact results by Bethe ansatz.\cite{JuettnerKluemper} We have chosen three chemical potentials corresponding in the low-temperature limit to a high, medium and low electron density (see Fig.~\ref{fig_density}). 
Obviously the particle density $n$ at a given chemical potential depends on temperature. Counting the degrees of freedom per lattice site immediately implies $n=2/3$ for any finite $\mu$ in the limit $T\rightarrow \infty$, whereas the density at finite temperature is given by $n = n(\mu,T)$.  
\begin{figure}
\includegraphics*[width=0.9\columnwidth]{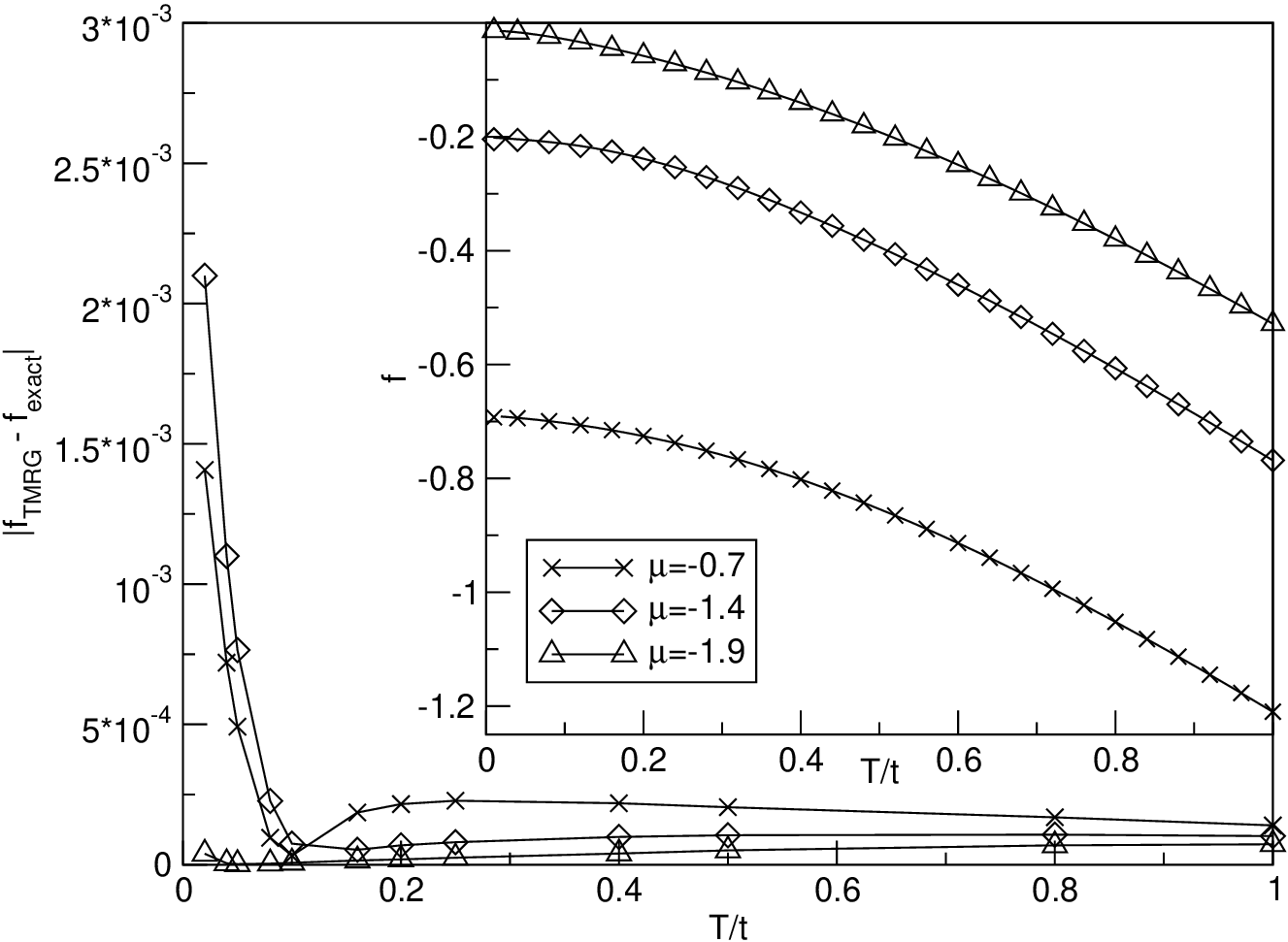}
\caption{Deviation of the free energies calculated by means of TMRG in comparison to the exact results for different temperatures. The lines are guide to the eye. The inset shows the Bethe ansatz results (symbols) and the TMRG results (lines).}
\label{fig1}
\end{figure}
For the same chemical potentials with $N=100$ states retained in the DMRG algorithm and $\epsilon=0.05$ the free energies and their accuracy are shown in Fig.~\ref{fig1} in comparison to the Bethe ansatz results by J\"uttner {\it et al.}~\cite{JuettnerKluemper} In all three cases the accuracy is of the order $10^{-4}$ if $T/t>0.1$. For lower temperatures $0.01 \leq T/t \leq 0.1$ the errors for the medium and high electron density grow up to $10^{-3}$, whereas the error in the low density case remains of the same order. More accurate results can be obtained if more states are retained or an extrapolation in the number of states $N$ and the small parameter $\epsilon$ is made. However, the obtained accuracy is sufficient for our purposes. From CFT it is known that the low-temperature asymptotics is given by
\begin{equation}
\label{conformal_free_energy}
f=e_0 -\frac{\pi}{6}\left(\frac{c_s}{v_s}+\frac{c_c}{v_c}\right)T^2
\end{equation}
where $v_{s,c}$ are the velocities of the spinon and holon excitations, respectively. Here the central charges  $c_{s,c}$ are equal to 1. 
According to Eq.~(\ref{conformal_free_energy}) we have fitted the numerical data and tried to determine the errors of the fit parameters by a variation of the fit region from $T/t\in [0.01:0.02]$ to $T/t\in [0.01:0.05]$. The estimates for the ground-state energies $e_0$ coincide with the exact results with deviations that are slightly larger than the errors calculated by the fit procedure (see table \ref{fit_values}). The remaining deviation is due to systematic errors. The parameter $1/v_s+1/v_c$ is difficult to obtain from such a fit, because the extent of the low-temperature region in which Eq.~(\ref{conformal_free_energy}) is valid is very small (see Bethe ansatz results in Fig.~\ref{fig2}) making it necessary to restrict the fit to the temperature region defined above. In such a small interval, however, a change of the parameter $1/v_s+1/v_c$ by a factor of 2 corresponds to deviations in the fitted free energy of the order $10^{-3}$ only, so that the TMRG data are not accurate enough to determine this parameter. However, it is possible to calculate the velocities from TMRG by using the results for the susceptibilities and correlation lengths to be discussed later on. We do not want to pursue this further, because the estimation of $v_s$ and $v_c$ is not our main goal.
\begin{table*}
\caption{Velocities $v_{s,c}$ and ground-state energies $e_0$ from the Bethe ansatz in comparison to values from a fit of the numerical data. The errorbars of $e_0$ and $1/v_s+1/v_c$ correspond to the described variation of the fit region, but the errors for $1/v_s+1/v_c$ are in fact much larger (see explanation in the main text).}
\label{fit_values}
\begin{tabular}[c] {|c||c|c|c|c|c|c|}
\hline \quad $\mu$ \quad\quad & \quad\quad $v_c$ \quad\quad & \quad $v_s$ \quad\quad & \quad $(1/v_s+1/v_c)^{\text{exact}}$ \quad\quad & \quad $(1/v_s+1/v_c)^{\text{fit}}$\quad\quad & \quad $e_0^{\text{exact}}$ \quad\quad & \quad $e_0^{\text{fit}}$\quad\quad \\
\hline\hline -0.7 & 0.526 & 2.778 & 2.261 & 3.90 ($\pm$ 0.61) & \quad -0.69114 \quad\quad &\quad -0.6908 $\pm$ 0.0002 \\
\hline -1.4 & 1.061 & 1.713 & 1.526 & 2.90 ($\pm$ 0.64) & \quad -0.20373 \quad \quad &\quad -0.2033 $\pm$ 0.0002 \\
\hline -1.9 & 0.579 & 0.657 & 3.249 & 3.66 ($\pm$ 0.14) &\quad -0.01320 \quad\quad&\quad -0.0133 $\pm$ 0.0001 \\
\hline
\end{tabular}
\end{table*}

To calculate the specific heat $c_n$, we first have to calculate the entropy $S$. This can be done by using the relation $S=(u-f)/T$, where the inner energy $u$ is directly calculated in the TMRG algorithm as a local expectation value or alternatively, by the numerical derivative $S=-\partial f/\partial T$. At first sight it seems better to calculate the entropy without using numerical derivatives, however, a local expectation value directly involves the eigenvectors of the QTM which are generally less accurate than the corresponding eigenvalues. By comparing both results with the exact ones we have convinced ourselves that the numerical derivative leads to more accurate results. Because in the calculations only the chemical potential $\mu$ but not the particle density $n$ can be fixed, we calculate the specific heat by using the thermodynamic relation
\begin{equation}
\label{specific_heat}
c_n=T\left(\frac{\partial S} {\partial T}\right)_n = T\left[\left(\frac{\partial S} {\partial T}\right)_\mu - \left(\frac{\partial n} {\partial T}\right)^2_\mu \left(\frac{\partial n} {\partial \mu}\right)^{-1}_T\right] \; .
\end{equation}
The second term is again evaluated using numerical derivatives. 
\begin{figure}[!ht]
\includegraphics*[width=0.9\columnwidth]{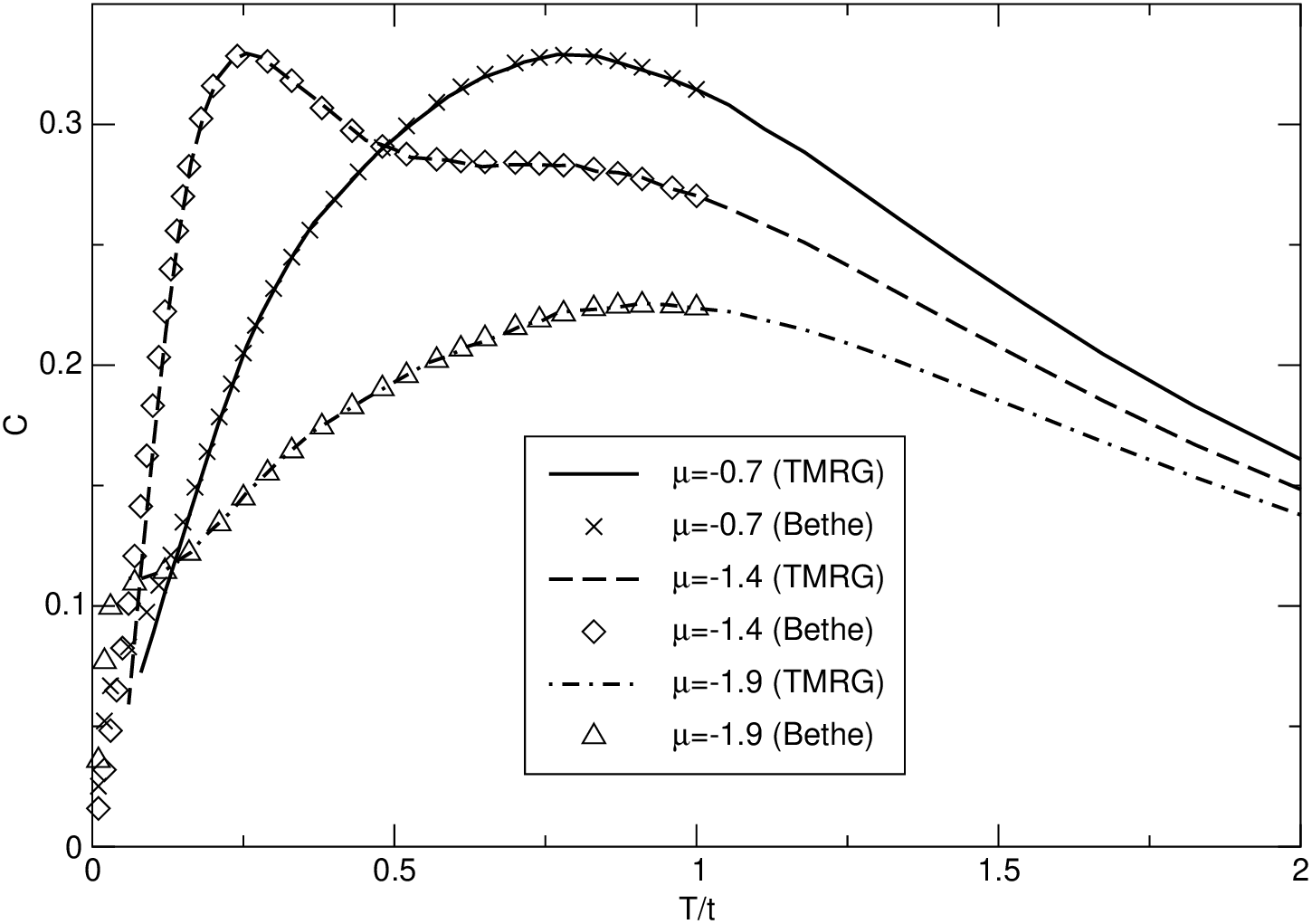}
\caption{Specific heats as calculated by means of TMRG and Eq.~(\ref{specific_heat}) (lines) and exact results by Bethe ansatz (symbols). Note the two peak structure caused by spin-charge separation as explained in the text.}
\label{fig2}
\end{figure}
This procedure enhances the numerical errors, but we are able to reproduce the
exact results within errors of the order $10^{-3}$ down to temperatures of
$T/t \approx 0.2$ as shown in Fig.~\ref{fig2}. 
The CFT result for the free energy, Eq.~(\ref{conformal_free_energy}), leads to a linear temperature dependence of the specific heat at low temperatures with a coefficient given by $\pi (1/v_c+1/v_s)/3$. The Bethe ansatz results clearly show this linear behavior at very low temperatures but within the TMRG this temperature region is not accessible because of the errors caused by the numerical derivatives used in Eq.~(\ref{specific_heat}). 

To calculate the spin susceptibility $\chi_s$, a small magnetic field $h=10^{-2}$ is applied. From the resulting magnetization $m$ the susceptibility at vanishing magnetic field is evaluated by
\begin{equation}
\label{spin_suscept}
\chi_s\big|_{h=0} = \frac{m}{h} \; .
\end{equation}
Similarly the charge susceptibility $\chi_c$ or compressibility is given by
\begin{equation}
\label{charge_suscept}
\chi_c = \frac{\partial n}{\partial \mu} \; ,
\end{equation}
where again numerical derivatives are used and the variation of $\mu$ is typically of the order $10^{-2}$.  
\begin{figure}[!ht]
\includegraphics*[width=0.9\columnwidth]{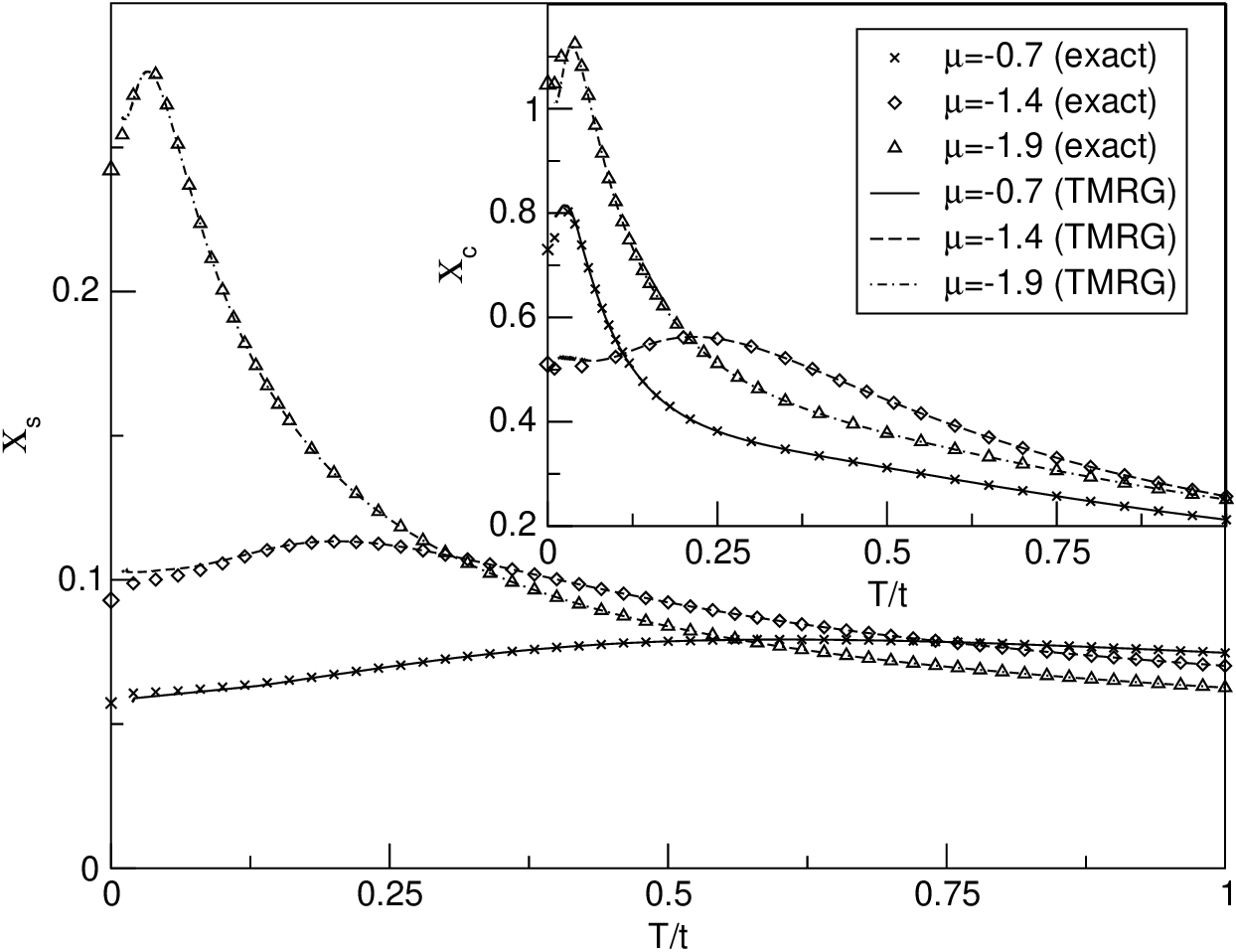}
\caption{The main figure shows the spin susceptibilities, the inset the charge susceptibilities (compressibilities), where again lines denote the numerical results and symbols the exact ones. The symbols at $T=0$ denote in both graphs the CFT results.}
\label{fig3}
\end{figure}
Here we are able to reproduce the exact results down to temperatures of $T/t=0.01$ as shown in Fig.~\ref{fig3}. The high temperature asymptotics of the spin susceptibility is given by the pure paramagnetic part $\chi_s(T\rightarrow \infty)\sim 2s(s+1)/9T$ what is easily understood when $\chi_s$ is expressed as a sum of two-point spin correlations. For $T=0$ the spin susceptibility has a finite value as expected from the linear dispersion of the spinon excitations. Within CFT the zero temperature value is given explicitly by $\chi_s = 1/(2\pi v_s)$ in good agreement with the numerical results. The maxima at finite temperatures are determined by the band structure.  
The charge susceptibility is given in the same way as sum over two-point density-density correlations leading to the high-temperature asymptotics $\chi_c\sim 2/9T$. For $T=0$ the charge susceptibility is again given by CFT as $\chi_c=\xi_c^2(Q)/(\pi v_c)$ with $\xi_c(Q)$ being the dressed charge. Note, that the charge susceptibility is diverging for $T\rightarrow 0$ in the two limiting cases $n\rightarrow 0$ and $n\rightarrow 1$, because $v_c\rightarrow 0$ in both cases. The spin susceptibility $\chi_s$ shows different behavior with
divergence for $T\to 0$ only in the limiting case $n\to 0$ as only
here $v_s\to 0$.

To summarize our numerical findings for $c$, $\chi_s$ and $\chi_c$ in
a qualitative manner we may use a picture of the ground state as a
``liquid'' consisting of bound singlet pairs of electrons. Above this
ground state there are two different elementary excitations. First, a
momentum transfer onto any individual pair is possible by keeping the
bound pair intact (holon excitation). Second, a breaking of any bound
pair into its components (electrons) with the two spin-1/2 objects
coupling either to $S=0$ or 1 (spinon excitation). Once the ``free''
electrons are produced by spinon excitations, they may acquire
momentum and energy individually which we refer to as incoherent
single particle motion.
Basically, any structure observed in the $T$-dependence of $c$,
$\chi_s$ and $\chi_c$ can be attributed to the saturation of a
particular type of excitation. In $\chi_s$ ($\chi_c$) we see a finite
temperature maximum at a temperature of the order of $v_s$ ($v_c$).
Note that the characteristic holon temperature is always
lower than or equal to the spinon temperature. This is rather
natural as the saturation of spinons is marked by the practical
absence of pairs such that saturation of holon excitations must have
occurred at lower or equal temperature. In principle, these structures
are also visible in the specific heat. In most cases, however, only
the spin-maximum is clearly noticeable. For generic densities the
small holon-maximum disappears in the low temperature regime of $c(T)$
with rather steep slope due to the spinon-excitations. At lower
densities the holon and spinon structures are located at similar
temperatures. In all cases, the higher temperature region is dominated
by a broad maximum due to incoherent single particle motion. The
corresponding energies are of the order of spinon excitations at
particle densities close to 1 in which case the structures merge.
\subsection{Correlation lengths}
For the correlation lengths of the supersymmetric t-J model no exact results are available. To obtain nevertheless a measure for the accuracy of the numerical data we have calculated the leading correlation length for a system of free spinless fermions (see appendix \ref{Appendix}). The comparison with the exact result shows that TMRG yields reliable results down to temperatures $T/t\sim 0.1$. Because the free fermion model is harder to tackle within the DMRG scheme due to the rapidly decaying spectrum of the reduced density matrix, we believe that for the t-J model a similar or even better accuracy is obtained. 
  
\begin{figure}[!ht]
\includegraphics*[width=0.9\columnwidth]{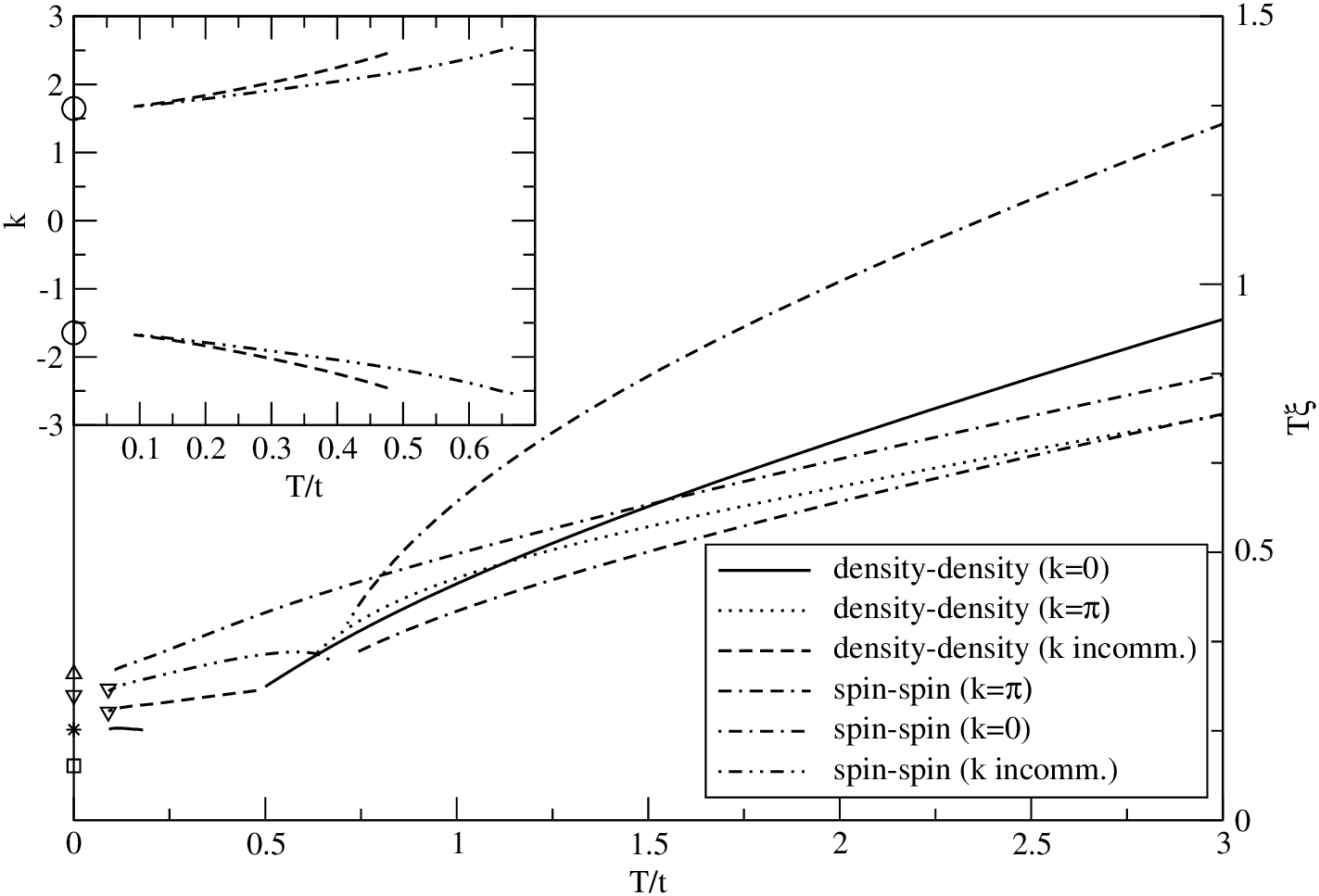}
\caption{Temperature dependence of the leading d-d and s-s CL's for $J/t=2.0$ and $\mu=-1.4$. The triangle up (star) gives the zero temperature result from CFT for the non-oscillating s-s (d-d) and the square that for the $4k_F$ part of the d-d. The triangle down at zero temperature denotes the CFT value for the $2k_F$ part of s-s and d-d, whereas the triangles down at $T/t=0.1$ are given by CFT plus logarithmic corrections as described in the text. The inset shows the wavevectors $k$ in the case of incommensurate oscillations. The circles denote the values for $k=\pm 2 k_F$ at zero temperature as expected from CFT.}
\label{fig4.1}
\end{figure}

We concentrate now on the calculation of correlation lengths for the supersymmetric t-J model to study the
crossover from the {\it high T lattice} into the {\it quantum critical
  regime} determined by $T\ll t$. In the high T lattice region we expect
non-universal properties dependent on the microscopic Hamiltonian $H$, whereas
the quantum critical regime should show universal TLL properties. Respecting the selection rules, the density-density (d-d CL's) and longitudinal spin-spin correlation lengths (s-s CL's) are in the block of the QTM with unchanged quantum numbers ($\Delta N_\uparrow = \Delta N_\downarrow = 0$). To distinguish between them, the matrixelement $M_\alpha$ in Eq.~(\ref{matrixel}) has to be calculated explicitly. With zero magnetic field and isotropic spin interactions, it turns out that $M_\alpha$ is either non-zero for the density or for the $S^z$ operator so that all d-d and s-s CL's are different from each other. This is a consequence of the $SU(2)$ spin symmetry. If a magnetic field is applied this symmetry is broken and less stringent selection rules apply such that all eigenvalues discussed above contribute to the s-s as well as to the d-d correlation function. On the other hand the singlet pair operator $P^s_i=c_{\uparrow,i+1}c_{\downarrow,i}$ changes the quantum numbers by $\Delta N_\uparrow = \Delta N_\downarrow = \pm 1$ and the eigenvalues corresponding to the singlet pair CL's are found in the block with these quantum numbers.  The leading d-d and s-s CL's (times temperature) and the corresponding wavevectors for $J/t=2.0$, $\mu=-1.4$ and zero magnetic field are shown in Fig.~\ref{fig4.1}. 
At high temperatures the largest s-s CL is given by a real negative eigenvalue
leading to $\pi$-oscillations. However, at a well defined crossover
temperature $T_c\approx 0.8$ a real, positive eigenvalue becomes largest,
which is associated with $k=0$. Regarding only the largest spin CL this means
that there is a non-analyticity at $T_c$. We want to point out that any
thermodynamic quantity derived from the free energy is an analytic function at
finite $T$. Phase transitions and corresponding singularities only occur at
$T=0$. However, quantities describing the {\it asymptotics of correlation
  functions} (i.e.~CL) may show non-analyticities even at finite
temperature. Such crossovers between CL's are characteristic for the
non-universal high T lattice regime, whereas no crossovers are expected in the
universal quantum critical regime described by CFT. The second largest spin CL is given at low temperatures by a pair of complex conjugated eigenvalues. Complex eigenvalues always appear in pairs, because the QTM is nonsymmetric but real and lead to incommensurate oscillations, i.e. $k\neq 0,\pi$. An incommensurate wavevector $k$ is depending on temperature as seen in the inset of Fig.~\ref{fig4.1}. The leading d-d CL shows no oscillations at high temperatures but is crossed at lower temperatures by a real, negative eigenvalue leading to $\pi$-oscillations before again the non-oscillating part dominates. Below $T_c\approx 0.5$ the largest d-d CL is given by a complex eigenvalue pair. The wavevector for this incommensurate part is again shown in the inset of Fig.~\ref{fig4.1}. \\
It is illuminating to regard first the d-d correlation for a system of free spinless fermions. For such a system the CL's can be easily calculated (see appendix \ref{Appendix}) and it turns out that a non-oscillating part and a part with $k=\pi$ arise, both with a CL given by $\xi^{-1}=2 \: \text{arcsinh} (\pi T)$. Therefore the CL's at high temperature behave as $\xi^{-1}\sim \ln(T)$ consistent with the numerical results. In the low temperature limit it follows that $\xi^{-1} \sim 1/(2\pi T)$ what is expected from CFT when the velocity $v$ and the scaling dimension $x$ are equal to $1$. The matrixelement for the CL is given by $M=2/(\pi^2+\beta^2)$ indicating that $M\rightarrow 0$ when $T\rightarrow 0$. This is reasonable, because at zero temperature the CL's diverge and the matrixelements have to approach zero so that the series in Eq.~(\ref{Corr-function}) is not diverging but sums up to an algebraic function. The numerical results also show that the matrixelements $M_\alpha \rightarrow 0$ if $T\rightarrow 0$.

To understand the behavior of the CL's at low temperatures in detail it is useful to regard the results from the Bethe ansatz and the finite-size scaling technique in CFT at zero temperature.\cite{KawakamiYang} In general, a two-point correlation function of the {\it scaling fields} $\phi_{\Delta^\pm}(r,\tau)$ with {\it conformal weights} $\Delta^\pm$ is given by
\begin{subequations}
\begin{eqnarray}
\label{CFT-corr}
&&\left<\phi_{\Delta^\pm}(r,\tau) \phi_{\Delta^\pm}(0,0) \right> \\
&=& \frac{\e^{\im(2\pi-2k_{F\uparrow}-2k_{F\downarrow})D_c x} \e^{\im(2\pi-2k_{F\uparrow})D_s x}}{(r-\im v_c\tau)^{2\Delta_c^+} (r+\im v_c\tau)^{2\Delta_c^-} (r-\im v_s\tau)^{2\Delta_s^+} (r+\im v_s\tau)^{2\Delta_s^-}} \nonumber
\end{eqnarray}
where the conformal weights are determined by quantum numbers of the elementary excitations. Considering here only the case of vanishing magnetic field and using the notation of Ref.~\onlinecite{KawakamiYang}, the conformal weights reduce to a simple form 
\begin{equation}
\label{conformal_weights1}
\Delta_c^\pm(\vec{I},\vec{D}) =\frac{1}{2}\left(\frac{I_c}{2\xi_c(Q)}\pm \xi_c(Q)\left(D_c+\frac{D_s}{2}\right)\right)^2+N_c^\pm
\end{equation}
\begin{equation}
\label{conformal_weights2}
\Delta_s^\pm(\vec{I},\vec{D}) =\frac{1}{4}\left(I_s-\frac{I_c}{2}\mp D_s\right)^2+N_s^\pm \: .
\end{equation}
$\xi_c(Q)$ is the dressed charge which can be calculated explicitly by solving an integral equation and $(I_c,I_s,D_c,D_s,N_c,N_s)$ are quantum numbers of the excitation. To be specific, $I_c$ counts the total number of holes, $I_s$ the number of holes with respect to the up spins, $D_c$ ($D_s$) the number of holons (spinons) being transfered from one Fermi point to the other and $N_c$ ($N_s$) the number of charge (spin) particle-hole excitations. The Fermi momentum $k_{F\uparrow(\downarrow)}$ for up (down) spin electrons is given by
\begin{equation}
\label{Fermi-mom}
k_{F\uparrow(\downarrow)} = \frac{\pi}{2}(n \pm 2m)
\end{equation}
\end{subequations}
where $m$ is the magnetization. If no magnetic field is applied then $k_{F\uparrow}=k_{F\downarrow}=k_F$. As shown in Fig.~\ref{fig_density} the density in the zero temperature limit for $\mu=-1.4$ is approximately $n_{T\rightarrow 0} \approx 0.524$ and therefore $k_F \approx 0.823$. In the inset of Fig.~\ref{fig4.1} the circles denote $k=\pm 2k_F$ at zero temperature indicating that the leading d-d and s-s CL's with incommensurate oscillations correspond to the $2k_F$-oscillating part.
From Eq.~(\ref{CFT-corr}) we derive 
\begin{subequations}
\begin{eqnarray}
\label{density-corr}
\left< n(r)n(0)\right> &=& \mbox{const.} + A_0 r^{-2} + A_2 r^{-\alpha_c} \cos(2k_Fr) \nonumber \\
&& + A_4 r^{-\beta_c}\cos(4k_Fr) 
\end{eqnarray}
for the equal-time d-d correlation, where $A_i$ are matrixelements. The non-oscillating part is due to the lowest particle-hole excitation $(0,0,0,0,1,0)$, the $2k_F$ part due to a $(0,0,\pm 1,\mp 1,0,0)$ excitation, while the $4k_F$ part arises from the excitation $(0,0,\pm 1,0,0,0)$. Thus, the critical exponents can be calculated from Eq.~(\ref{conformal_weights1},\ref{conformal_weights2}) leading to 
\begin{equation}
\label{crit-exp}
\alpha_c = 1+\frac{\xi_c^2(Q)}{2}     \quad\quad , \quad\quad    \beta_c = 2 \xi_c^2(Q)    \: .
\end{equation}
\end{subequations}
The equal-time s-s correlation has the same form as Eq.~(\ref{density-corr}), but the constant as well as the $4k_F$ part are absent and the matrixelements are different. The critical exponent of the $2k_F$ part is the same but the corresponding excitation is now $(0,0,0,\pm 1,0,0)$.\footnote{The excitations $(0,0,\pm 1,\mp 1,0,0)$ and $(0,0,0,\pm 1,0,0)$ lead to the same critical exponent but they can be distinguished by applying a magnetic field because the first shows $2k_{F\downarrow}$-oscillations whereas the second oscillates with $2k_{F\uparrow}$.}
As shown in Ref.~\onlinecite{KawakamiYang} the dressed charge $\xi_c(Q)$ varies between $\sqrt{2}$ for $n=0$ and $1$ at half-filling. Thus, $\alpha_c$ is always the smallest critical exponent and therefore the $2k_F$ parts of the correlation functions dominate at zero temperature. 

The results for the correlation functions shown here are absolutely consistent with TLL theory when the identification
\begin{equation}
\label{K_rho}
K_\rho = \frac{\xi_c^2(Q)}{2}
\end{equation} 
is used. For the Tomonaga-Luttinger model it is also possible to calculate multiplicative logarithmic corrections to the algebraic terms of Eq.~(\ref{density-corr}) which are not directly accessible within Bethe ansatz and CFT. It turns out that the $2k_F$ parts of the d-d and the s-s correlation have different logarithmic corrections given by $\ln^{-3/2}r$ ($\ln^{1/2}r$) for the d-d (s-s) correlation.\cite{GiamarchiSchulz}

The long-distance asymptotics of correlation functions at small finite temperature can still be obtained from conformal invariance by the usual mapping of the complex plane onto a strip of width $1/T$ with periodic boundary conditions. Eq.~(\ref{CFT-corr}) is then replaced by
\begin{widetext}
\begin{eqnarray}
\label{conf-mapping}
\e^{\im(2\pi-2k_{F\uparrow}-2k_{F\downarrow})D_c x} \e^{\im(2\pi-2k_{F\uparrow})D_s x} && \left(\frac{\pi T}{v_c\sinh(\pi T(x-\im v_c\tau)/v_c)}\right)^{2\Delta_c^+}  \left(\frac{\pi T}{v_c\sinh(\pi T(x+\im v_c\tau)/v_c)}\right)^{2\Delta_c^-} \nonumber \\ 
&\times& \left(\frac{\pi T}{v_s\sinh(\pi T(x-\im v_s\tau)/v_s)}\right)^{2\Delta_s^+}  \left(\frac{\pi T}{v_s\sinh(\pi T(x+\im v_s\tau)/v_s)}\right)^{2\Delta_s^-} \: .
\end{eqnarray}
\end{widetext}
Thus, all CL's diverge in the low temperature limit as
\begin{equation}
\label{T_dependence}
\xi = \frac{1}{2\pi T \left(\frac{x_c}{v_c}+\frac{x_s}{v_s}\right)} = \frac{\gamma}{T}
\end{equation}
where the coefficient $\gamma = v_s/2\pi$ ($\gamma = v_c/2\pi$) for the non-oscillating part of the s-s (d-d) CL,
\begin{subequations}
\begin{equation}
\label{gamma_2k_F}
\gamma = \frac{2 v_c}{\pi \left(2\frac{v_c}{v_s}+\xi_c^2(Q)\right)}
\end{equation}
for both $2k_F$ parts whereas 
\begin{equation}
\label{gamma_4k_F}
\gamma = \frac{v_c}{2 \pi \xi_c^2(Q)}
\end{equation}
\end{subequations}
for the $4k_F$ part of the d-d CL. Contrary to zero temperature the answer to the question which part dominates depends not only on the scaling dimensions $x_{s,c}$ but also on the velocities $v_{s,c}$! In Fig.~\ref{fig4.1} the values for these $\gamma$'s are denoted by symbols at $T=0$ showing good agreement for the non-oscillating parts between the numerics and the CFT results. The $4k_F$-oscillating d-d CL is so small that it is difficult to obtain numerically. From Eq.~(\ref{gamma_2k_F}) the $2k_F$ s-s and d-d CL's are expected to be equal in the low-temperature limit. However, the numerical result shows that they are well separated even at the lowest accessible temperatures. This is a consequence of the different logarithmic corrections stated above. Quantitatively we can regard a multiplicative logarithmic term $\ln^\alpha r$ as an effective, distance dependent correction of the scaling dimension:
\begin{equation}
\label{log_corr}
x' = x - \frac{1}{2}\frac{\ln(\ln^\alpha r)}{\ln r}
\end{equation} 
The relevant length scale is given by the correlation length at the considered
temperature, $r \approx \xi(T)$. Using this correction together with Eq.~(\ref{T_dependence}) leads to splitting of the $2k_F$ s-s and d-d CL at finite temperature and to an excellent agreement with the numerical results (see triangle downs in Fig~\ref{fig4.1}).

\begin{figure}[!ht]
\includegraphics*[width=0.9\columnwidth]{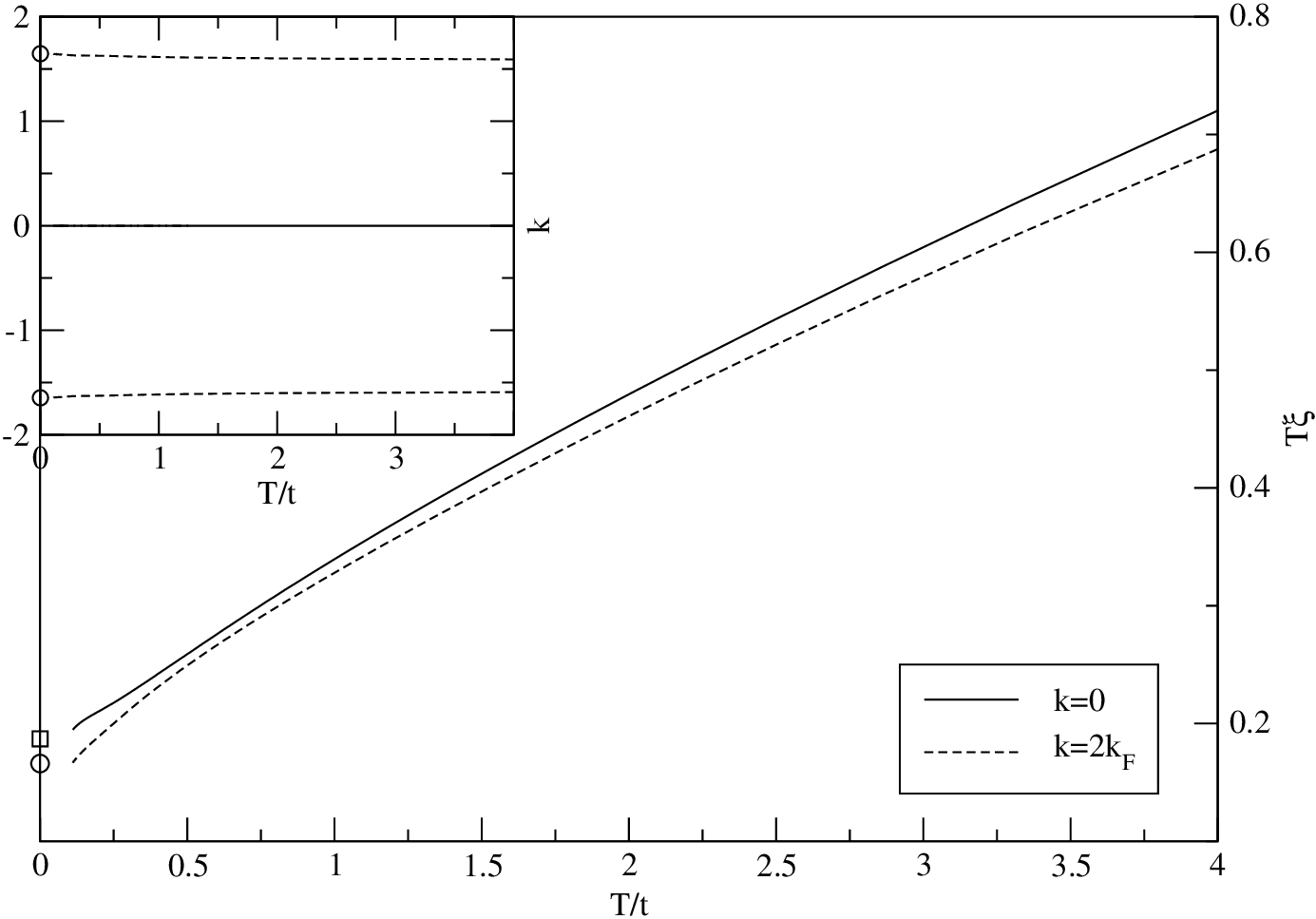}
\caption{Leading singlet pair correlation lengths for $J/t=2.0$ and $\mu=-1.4$. The square at $T=0$ denotes the CFT result for the non-oscillating part, the circle that for the $2k_F$ part. The inset shows the corresponding wavevectors.}
\label{fig_pair-corr}
\end{figure}

Next, we regard the singlet pair CL's shown in Fig.~\ref{fig_pair-corr}. 
The CL leading over the entire temperature range is non-oscillating, whereas the next-leading shows incommensurate oscillations. Again the zero temperature result from Eq.~(\ref{Fermi-mom}) identifies this contribution as the $2k_F$ part. 
The form of the algebraically decaying singlet pair correlation $P_s(r,0)$ at zero temperature is given by CFT
\begin{subequations}
\begin{equation}
\label{CFT_pair-corr}
P_s(r,0) = C_0 r^{-\delta_p} + C_2 r^{-\epsilon_p} \cos(2k_Fr)  \: .
\end{equation}
The non-oscillating term is due to the excitation $(2,1,\pm 1/2,\mp 1,0,0)$ leading to an exponent
\begin{equation}
\label{pair-exp}
\delta_p = 1 + \frac{2}{\xi_c^2(Q)}
\end{equation}
and the $2k_F$ part due to the excitation $(2,1,\pm 1/2,0,0,0)$ leading to
\begin{equation}
\label{pair-exp2}
\epsilon_p = \frac{2}{\xi_c^2(Q)} + \frac{\xi_c^2(Q)}{2} \: . 
\end{equation}
\end{subequations}
Thus, the coefficient $\gamma$ in Eq.~(\ref{T_dependence}) is now
\begin{subequations}
\begin{equation}
\label{gamma-pair}
\gamma = \frac{v_c}{\pi \left(\frac{2}{\xi_c^2(Q)}+\frac{v_c}{v_s}\right)} 
\end{equation}
for the non-oscillating part, whereas
\begin{equation}
\label{gamma-pair2}
\gamma = \frac{v_c}{\pi\left(\frac{2}{\xi_c^2(Q)}+\frac{\xi_c^2(Q)}{2}\right)}  
\end{equation}
\end{subequations}
for the $2k_F$ part. The value for the non-oscillating part is again in good agreement with the numerical results whereas the numerical accessible temperature range is not sufficient to compare the $2k_F$ part with CFT (see symbols in Fig.~\ref{fig_pair-corr}).

\begin{figure}[!ht]
\includegraphics*[width=0.9\columnwidth]{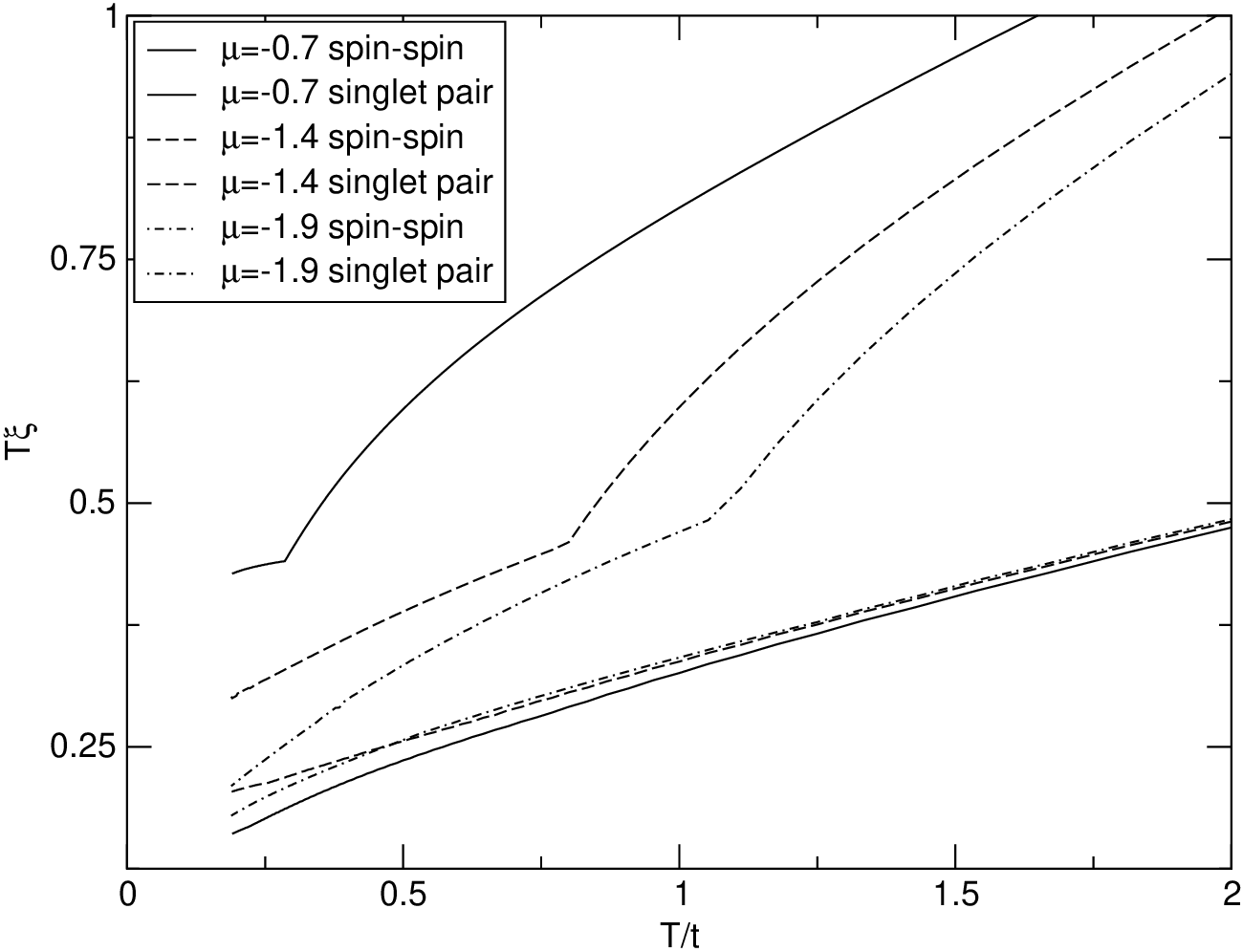}
\caption{The three upper curves show the spin correlation lengths, the other three the singlet pair-pair correlation lengths.}
\label{fig4}
\end{figure}

To illustrate the dependence of the leading s-s and singlet-pair CL on electron density, there are plots for three different chemical potentials in Fig.~\ref{fig4}. 
The s-s CL is largest for the Heisenberg case ($n=1$) and gets suppressed, as expected, with an increasing concentration of holes. On the other hand the singlet-pair CL is nearly independent of the particle density. 
The s-s CL shows a non-analyticity at a temperature $T_c$, where the $k=\pi$ part is crossed by the non-oscillating part. With increasing particle density this non-analyticity is shifted to lower temperatures and disappears for chemical potentials $\mu > -0.47$. Therfore no singularity is visible in the Heisenberg limit of the t-J model. $T_c$ as a function of the chemical potential shows an algebraic behavior. 
A very similar phenomena is described in Ref.~\onlinecite{KluemperScheeren} where the s-s CL of the $1/2$-XXZ chain in a magnetic field is investigated. In this case the magnetic field plays the same role as the chemical potential here. However, the singularity there is associated with a crossover from commensurate oscillations at $T>T_c$ to incommensurate oscillations at $T<T_c$. Recently, temperature dependent CL's have also been studied analytically in a generalized Hubbard model related to the supersymmetric t-J model.\cite{SakaiKluemper}
\subsection{Static correlation functions}
Within the TMRG imaginary time correlation functions are easy to calculate.\cite{Peschel} The fatal point is that the analytical continuation of the imaginary time results to real times is an ill-posed problem. We therefore want to restrict ourselves to the calculation of static correlation functions defined by
\begin{equation}
\label{def_static}
G(r,z=0) = \int_0^\beta d\tau \; G(r,\tau)
\end{equation}
where $G(r,\tau)$ is the correlation function for distance $r$ and imaginary time $\tau$. Note, that the static correlation function is calculated to higher accuracy than the equal-time correlation. Nevertheless the spatial CL's are identical in both cases except for the trivial correlations caused by particle-hole excitations which have vanishing matrixelements in the static case (see Eq.~\ref{conf-mapping}). The static correlation function can be expressed through the largest eigenvalue and corresponding eigenvectors of the QTM:
\begin{equation}
\label{static2}
G(r,z=0) = \frac{\epsilon}{M \Lambda_0^{r+1}} \langle \Psi^L_0 | \tilde{T}_M T_M^{r-1} \tilde{T}_M | \Psi^R_0 \rangle
\end{equation}
for distances $r\geq 1$ with the definition
\begin{equation}
\label{static2.1}
\tilde{T}_M = \sum_{k=0}^M T_M(A_{\epsilon \cdot k})
\end{equation}
where $T_M(A_{\epsilon \cdot k})$ denotes the usual transfer matrix $T_M$ where the considered operator $A$ is added at imaginary time position $\tau=\epsilon \cdot k$ and
\begin{equation}
\label{static3}
G(r=0,z=0) = \frac{\epsilon}{\Lambda_0} \langle \Psi^L_0 | \hat{T}_M | \Psi^R_0 \rangle 
\end{equation}
with
\begin{equation}
\label{static3.1}
\hat{T}_M = \sum_{k=0}^M T_M(A_0,A_{\epsilon \cdot k}) \; .
\end{equation}
These equations are valid within the alternative mapping to the classical model with alternating rows\cite{SirkerKluemper2} where a local transfer matrix covers only one site (see Sec.~\ref{Intro}). Note, that these formulas become more complicated if the usual decomposition of the Hamiltonian into even and odd parts is applied, because in this case even and odd distances as well as distance $r=1$ have to be treated separately. To check the accuracy of the TMRG algorithm we have calculated the static density-density correlation function for free spinless fermions (see appendix \ref{Appendix}) showing that the TMRG results are reliable even at short distances $r$.  
\begin{figure}[!ht]
\includegraphics*[width=0.9\columnwidth]{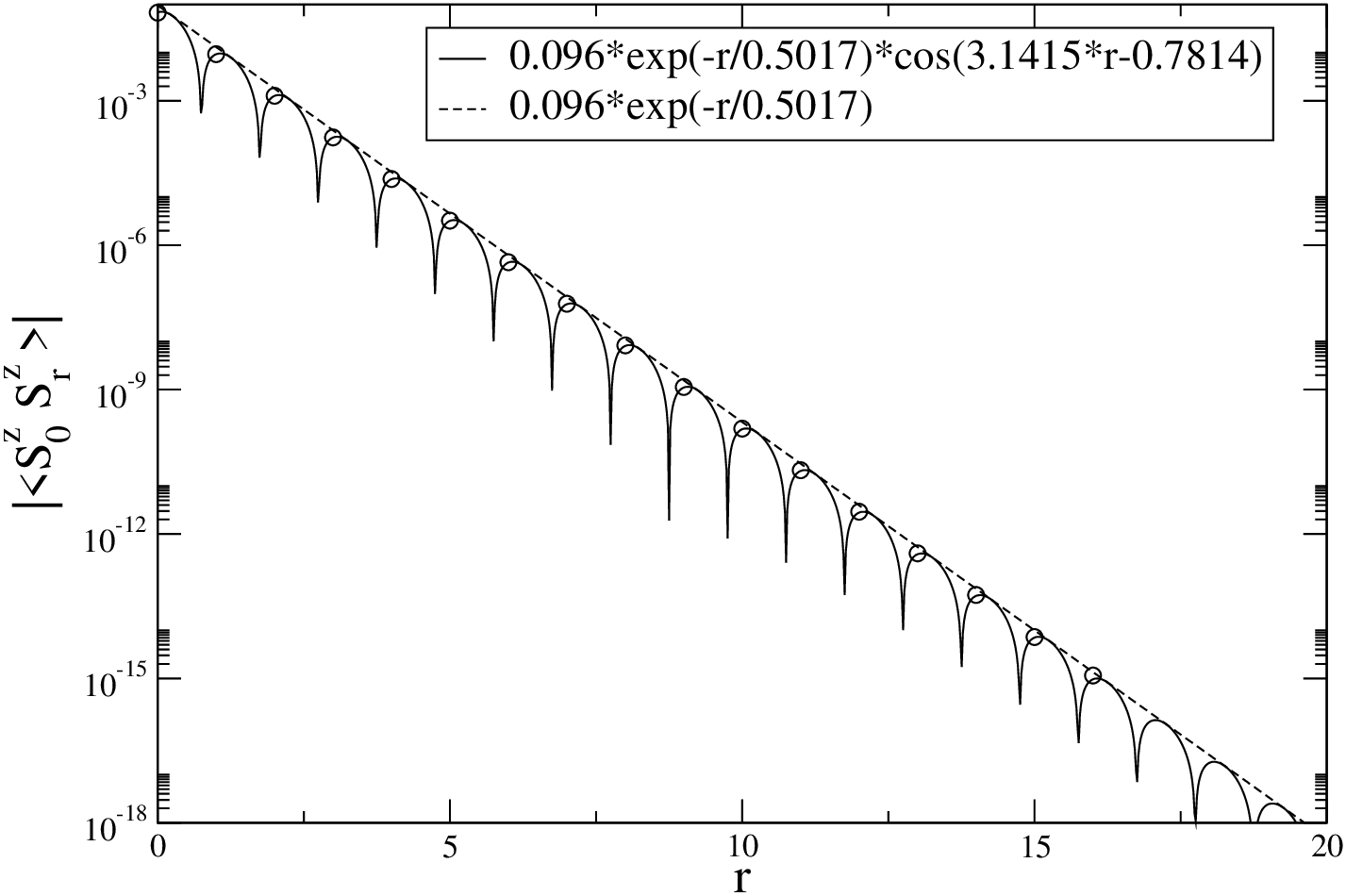}
\caption{Static longitudinal s-s correlation function at $T=2.0$ showing $\pi$-oscillations. The dotted line denotes an envelope corresponding to the exponential decay. The solid line is a guide to the eye.}
\label{stat_corr1}
\end{figure}
In Fig.~\ref{stat_corr1} (Fig.~\ref{stat_corr2}) the static longitudinal s-s correlation function for $J/t=2.0$, $\mu=-1.4$ and $T=2.0$ ($T=0.1$) is shown. From Fig.~\ref{fig4.1} we expect $\pi$-oscillations of the correlation function for $T=2.0$ and $2k_F$-oscillations for $T=0.1$ because the non-oscillating part does not show up in a static correlation function as explained before. 
\begin{figure}[!ht]
\includegraphics*[width=0.9\columnwidth]{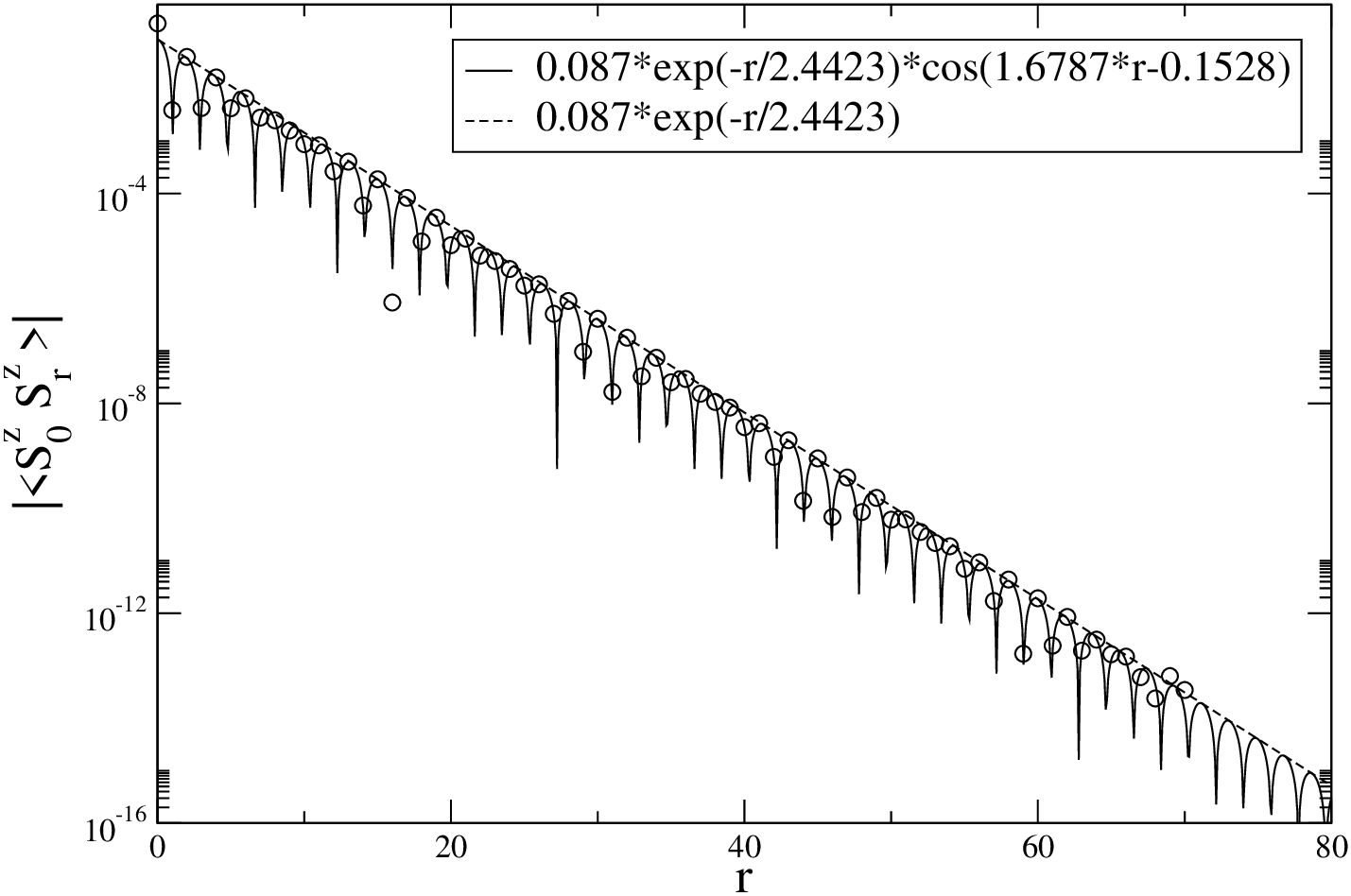}
\caption{Static longitudinal s-s correlation function at $T=0.1$ showing incommensurate oscillations. The dotted line denotes an envelope corresponding to the exponential decay. The solid line is a guide to the eye.}
\label{stat_corr2}
\end{figure}
By using $\langle S^z_0 S^z_r \rangle = A \exp(-r/\xi)\cos(kr+\delta)$ as a
fit function we have estimated the CL's as well as the wavevectors for both
temperatures directly from the correlation functions (see legends in
Fig.~\ref{stat_corr1} and Fig.~\ref{stat_corr2}). In both cases the results
coincide within errors of the order $10^{-4}$ with that obtained by using
directly the eigenvalues of the QTM. The perfect coincidence of the CL's estimated directly from the eigenvalues and that from a fit of the correlation function is not astonishing. In both cases the same QTM is used in the numerics where several eigenvalues are calculated directly using diagonalization routines in the first case whereas only the largest eigenvalue and the corresponding eigenvectors are needed to calculate the correlation function by some time-consuming matrix-vector multiplications in the second case. However, it provides a good consistency check of the numerics.   
\section{The physically relevant region}
\label{PhysRel}
As mentioned in the introduction the t-J model is a simplification of the Hubbard model obtained in the limit $U\gg t$. Thus, one is especially interested in the case where the spin exchange coupling $J=2t^2/|U|$ is smaller than the hopping $t$. The TMRG is applicable for any value of $J/t$ with almost the same accuracy and our algorithm remains stable even in the limiting cases $J/t \ll 1$ or $J/t \gg 1$. Problems as reported by Ammon {\it et al.}~\cite{RiceTroyer} never occurred in our algorithm. To be specific, we never observed that the algorithm breaks down after performing only a few or more RG-steps although we have retained up to 240 states to calculate the correlation lengths accurately. In the following calculations we set $J/t=0.35$, a value often used in the literature.
\begin{figure}[!ht] 
\includegraphics*[width=0.9\columnwidth]{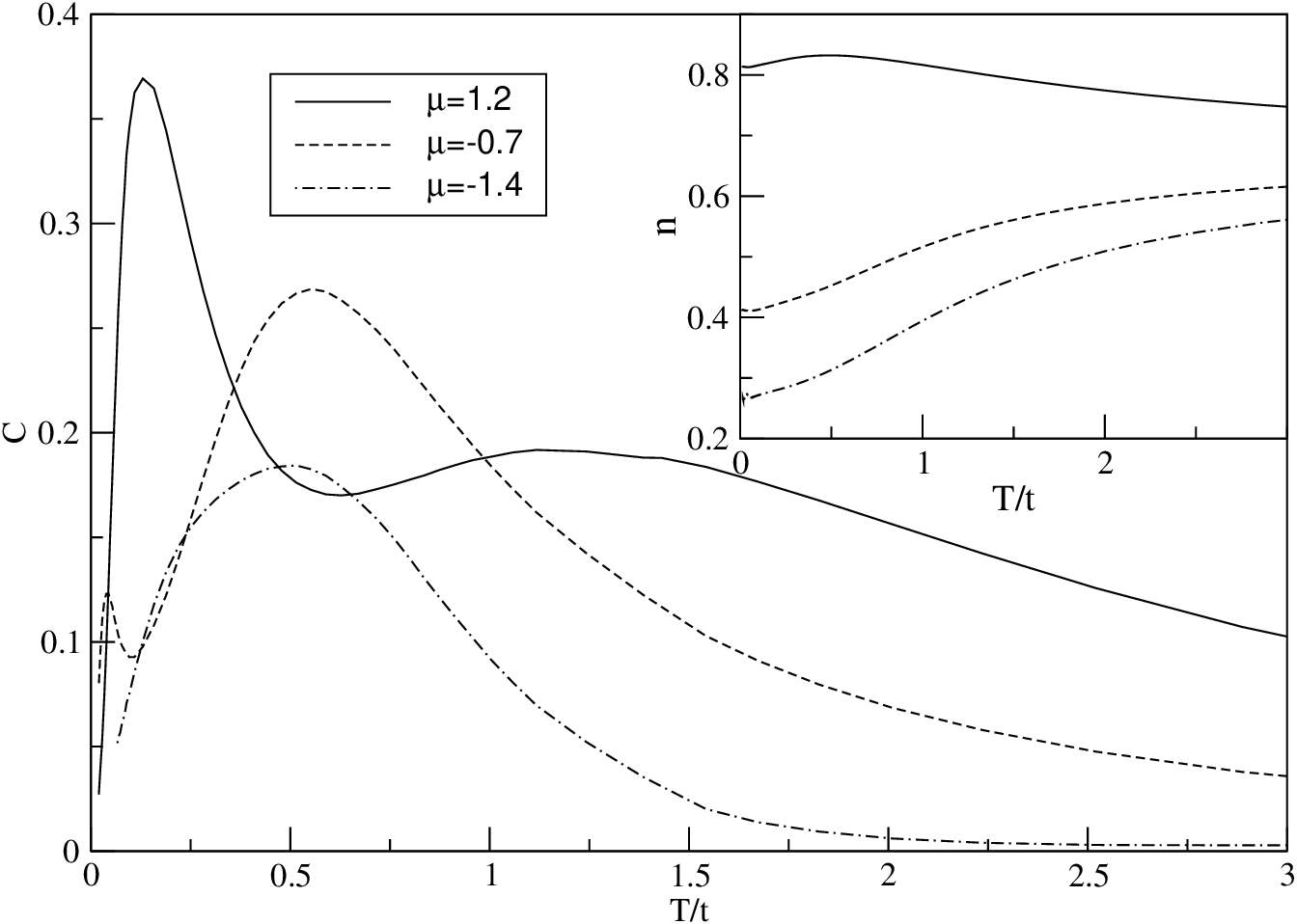}
\caption{Specific heat for $J/t=0.35$ and three different chemical potentials corresponding in the low temperature limit to a high ($\mu=1.2$), medium ($\mu=-0.7$) and low ($\mu=-1.4$) electron density. Note again the two-peak structure due to spin-charge separation. The inset shows the temperature dependence of the densities $n(\mu,T)$ for the same chemical potentials.}
\label{fig_spec_Heat}
\end{figure}

\begin{figure}[!ht] 
\includegraphics*[width=0.9\columnwidth]{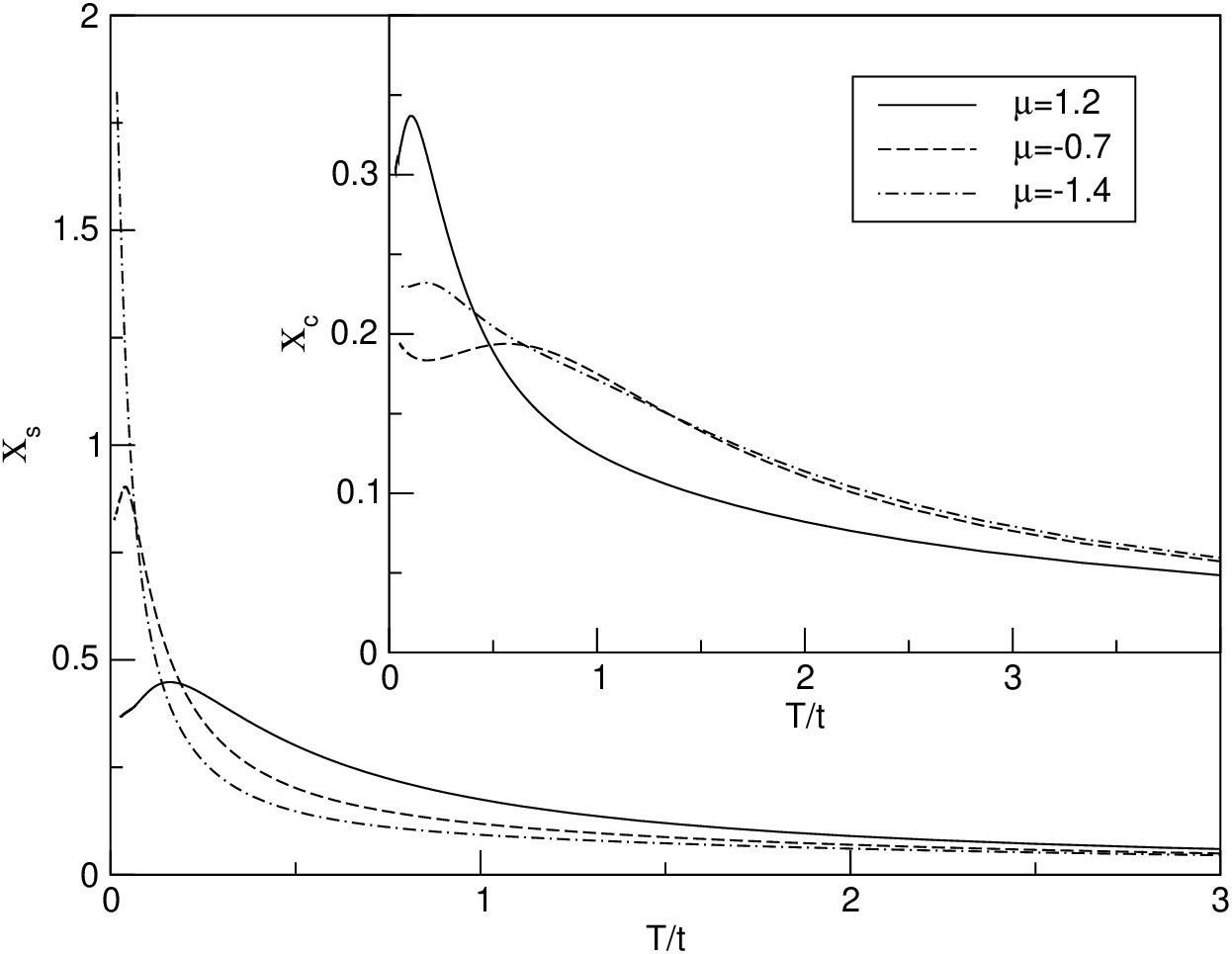}
\caption{Susceptibilities and compressibilities for the three different chemical potentials and $J/t=0.35$.}
\label{fig9}
\end{figure}

When regarding the specific heat shown in Fig.~\ref{fig_spec_Heat} a two-peak structure is again obvious in the high and medium density case and there seems to be a linear regime consistent with TLL theory, although we have noticed in the supersymmetric case that numerical errors grow up for low temperatures. The structure of the spin and charge susceptibilities (see Fig.~\ref{fig9}) looks also rather similar to the supersymmetric case.
The arguments given for the high and low temperature regime of the susceptibilities are also applicable here. It is easy to understand that the absolute values of the spin susceptibility are now larger, because the weaker antiferromagnetic coupling $J$ suppresses less efficiently a ferromagnetic alignment of spins. 
In principle, the qualitative discussion of $c$, $\chi_s$ and $\chi_c$ is very similar to that of the
integrable case. There are, however, two modifications. First, in the
charge susceptibility for low densities we observe two maxima. We do
not have any simple explanation of this fact. Note that a similar
observation has been made for the Hubbard chain. Second, due to the
different energy scales of spinon and incoherent single particle
motion at any density there are two maxima in $c(T)$ even at densities
close to 1.
\begin{figure}[!ht] 
\includegraphics*[width=0.9\columnwidth]{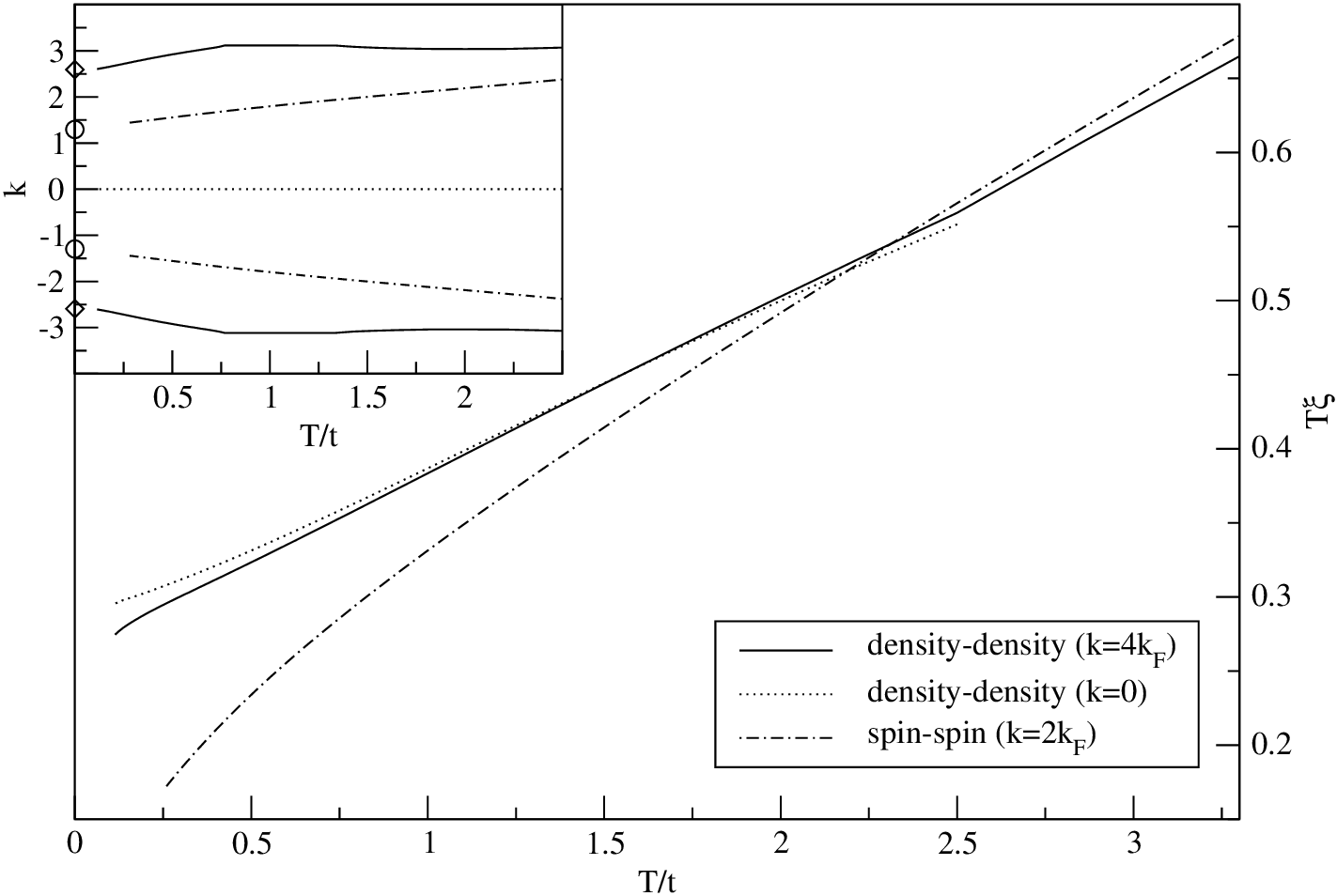}
\caption{S-s and d-d CL's for $J/t=0.35$ and $\mu=-0.7$. The inset shows the corresponding wavevectors. The circles (diamonds) denote the CFT result for $\pm 2k_F$ ($\pm 4k_F$).}
\label{fig_TLL_Corrlength}
\end{figure}

We also calculated again several s-s and d-d CL's as shown in Fig.~\ref{fig_TLL_Corrlength}. The CL's are now often difficult to distinguish and the crossing of CL's is ``smoother''  and shifted to higher temperatures when compared with the supersymmetric case. This clearly shows that the spin-exchange interaction is responsible for the various crossovers of CL's and in Appendix \ref{Appendix} we show that all such crossovers vanish for free spinless fermions. It is also striking that incommensurate oscillations are now present even at high temperatures. The leading s-s CL shows oscillations which are identified by Eq.~(\ref{Fermi-mom}) as $2k_F$ whereas one of the d-d CL shows $4k_F$ oscillations. The other d-d CL from Fig.~(\ref{fig_TLL_Corrlength}) is non-oscillating and leading in the low-temperature limit. These kind of oscillations are expected from TLL theory (see Eq.~(\ref{density-corr})) and we also note that the CL's seem to diverge again as $\xi\sim 1/T$ for low temperatures. This provides additional evidence that the t-J model at $J/t=0.35$ belongs to the same universality class (TLL) as the supersymmetric model.
\section{Phase separation}
\label{Phasesep}
At $J/t$ large, the attractive Heisenberg interaction in Eq.~(\ref{t-J_Ham}) dominates the kinetic energy term. In a canonical ensemble the model therefore phase separates into a high density and a low density region in order to optimize the Heisenberg energy. In our grand-canonical description of the model, however, there remains a competition between the chemical potential term and the Heisenberg exchange energy. If we ignore the kinetic energy completely - what is exact if total phase separation occurs - a simple picture evolves. Because the ground state energy of a $s=1/2$-Heisenberg chain per particle is given by $-J\ln2$, we expect in the limit $T\rightarrow 0$ an empty state if $\mu<-J\ln2$ and a state with $n=1$ if $\mu>-J\ln2$. Therefore phase separation can only be present at one special point characterized by $\mu\approx -J\ln 2$.
\begin{figure}[!ht] 
\subfigure[$J/t=3.1$]{
\includegraphics*[height=0.18\textheight,width=0.45\columnwidth]{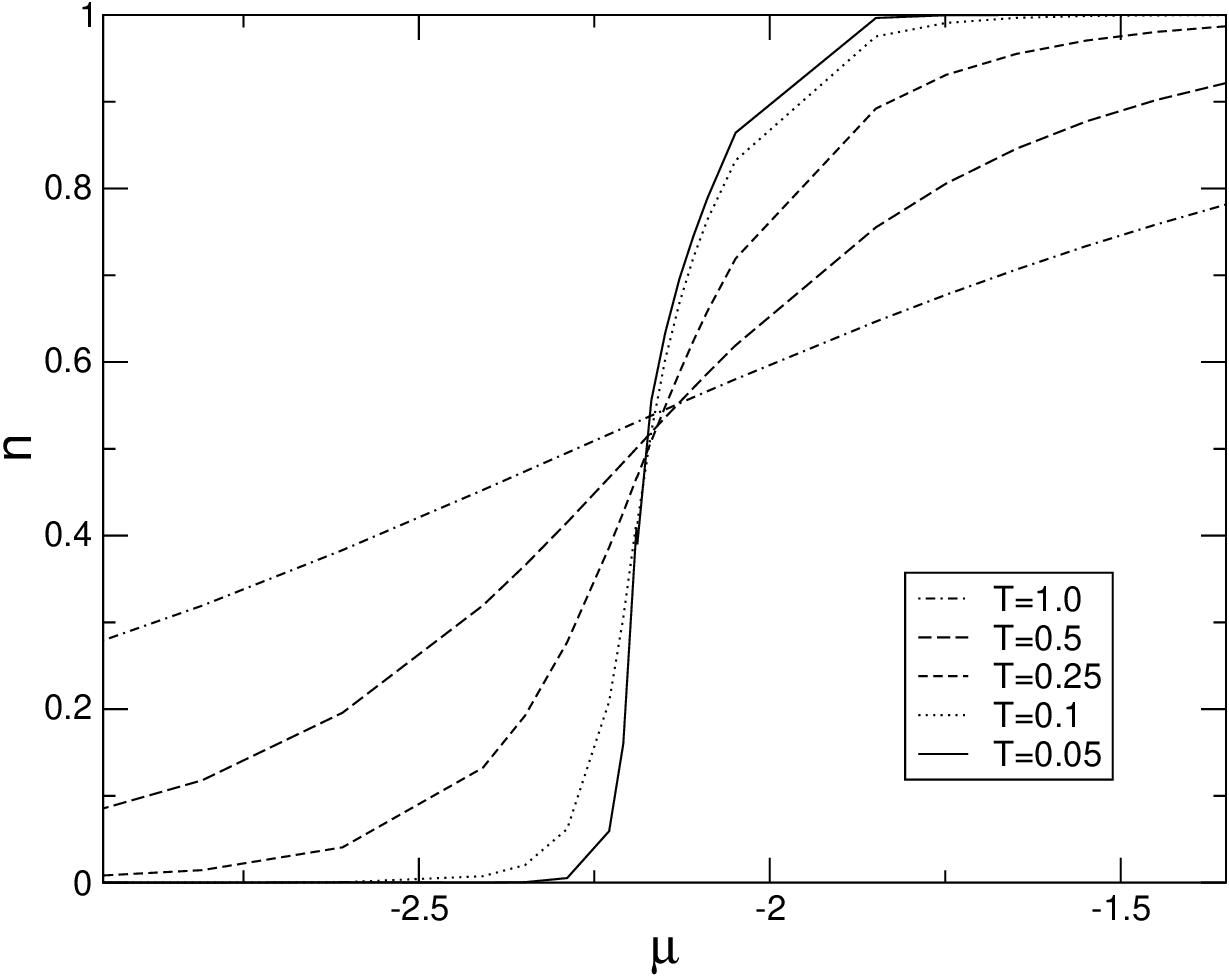}
\label{fig5}}
\hspace{0.1cm}
\subfigure[$J/t=3.3$]{
\includegraphics*[height=0.18\textheight,width=0.45\columnwidth]{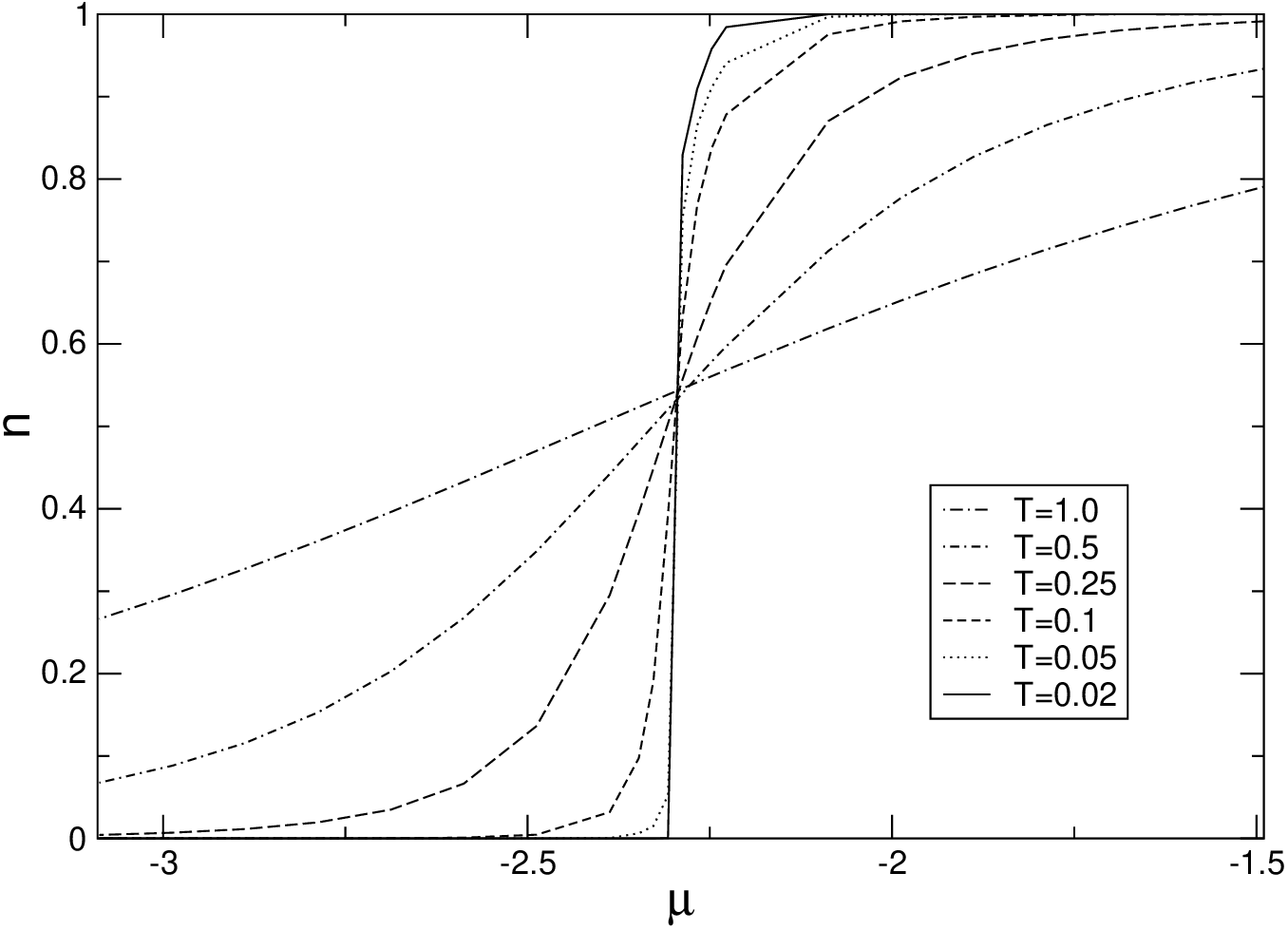}
\label{fig6}}

\subfigure[$J/t=3.5$]{
\includegraphics*[width=0.9\columnwidth]{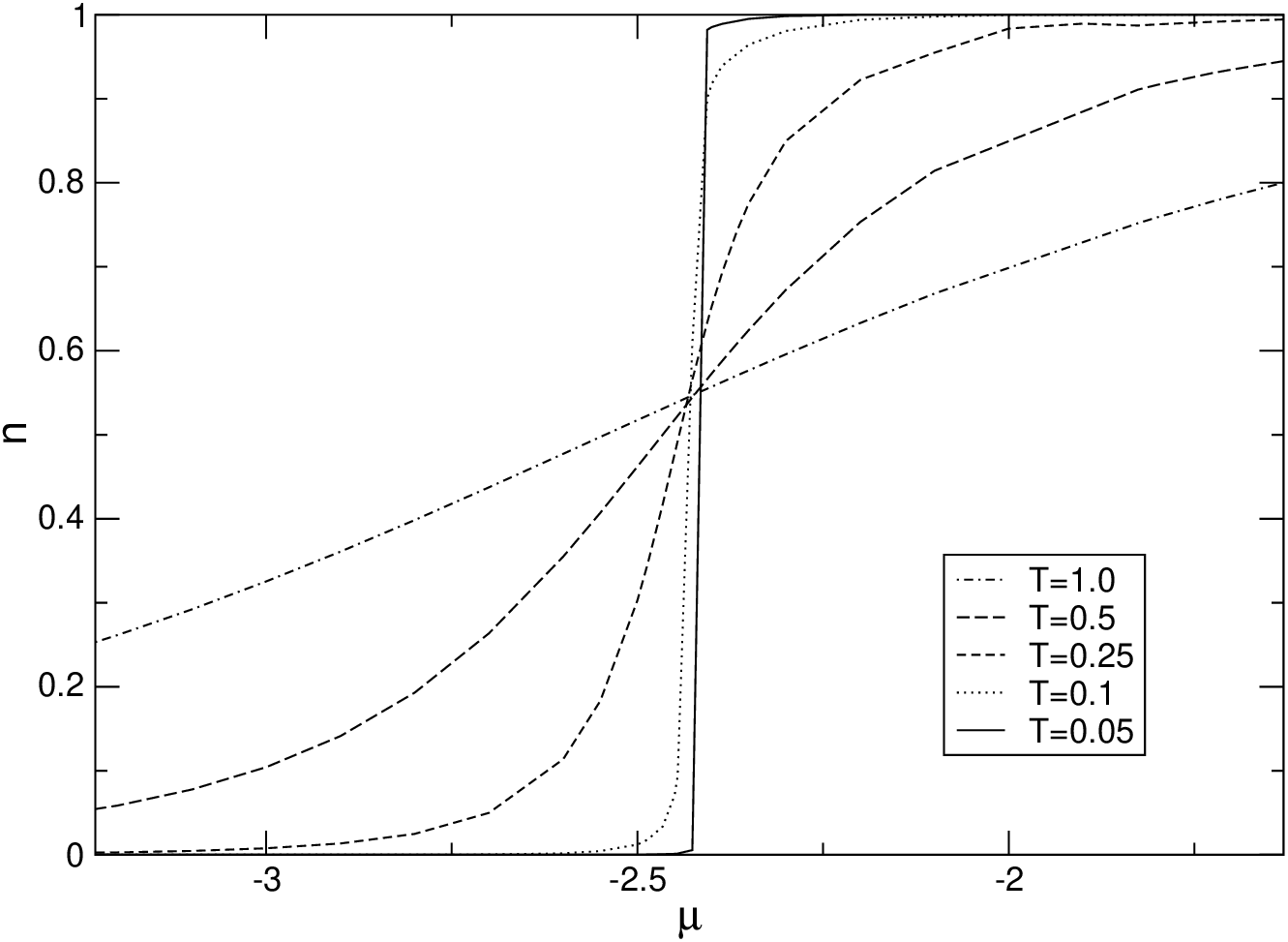}
\label{fig7}}
\caption{The figures show the density as a function of the chemical potential for fixed temperatures $T$ in units of $t$. Note that the lines connect a finite number of data points.}
\end{figure}
In the Figures \ref{fig5}, \ref{fig6}, \ref{fig7} the density at a constant
temperature is shown as a function of the chemical potential for three
different parameters $J/t$. Fig.~\ref{fig5} shows that at $J/t=3.1$ the
compressibility $\partial n/\partial \mu$ in the limit $T\rightarrow 0$ is not
diverging, indicating that the phase separated region is not yet reached. At
$J/t=3.3$ (Fig.~\ref{fig6}) the density jumps from $n=0$ to $n\approx 0.8$ at
$\mu\approx -J\ln2$ in the limit $T\rightarrow 0$ as expected for the phase separated region. The phase
separation is here between the empty and an electron rich state. At $J/t=3.5$
we find full phase separation, indicated by the jump of the density from 0 to
1 for $T\rightarrow 0$ (see Fig.~\ref{fig7}). Our calculations confirm that
the fully phase-separated state is destroyed by introducing holes into the
Heisenberg chain island as stated by Ogata {\it et al.}~\cite{OgataLuchini}
\section{Luther-Emery Phase}
\label{Luther-Emery}
As already mentioned in the introduction a phase with a spin gap is expected at least for low densities and values of $J/t$ close to phase separation. Because we expect the spin gap to be caused by a marginal operator leading to an exponentially small gap, the TMRG is not suited to determine the phase boundaries. However, in regions where the spin gap is larger than the lowest accessible temperature the TMRG can show the existence of such a phase without using any additional assumptions. A spin gap is directly visible in the spin susceptibility going to 0 for $T\rightarrow 0$. 
\begin{figure}[!ht] 
\includegraphics*[width=0.9\columnwidth]{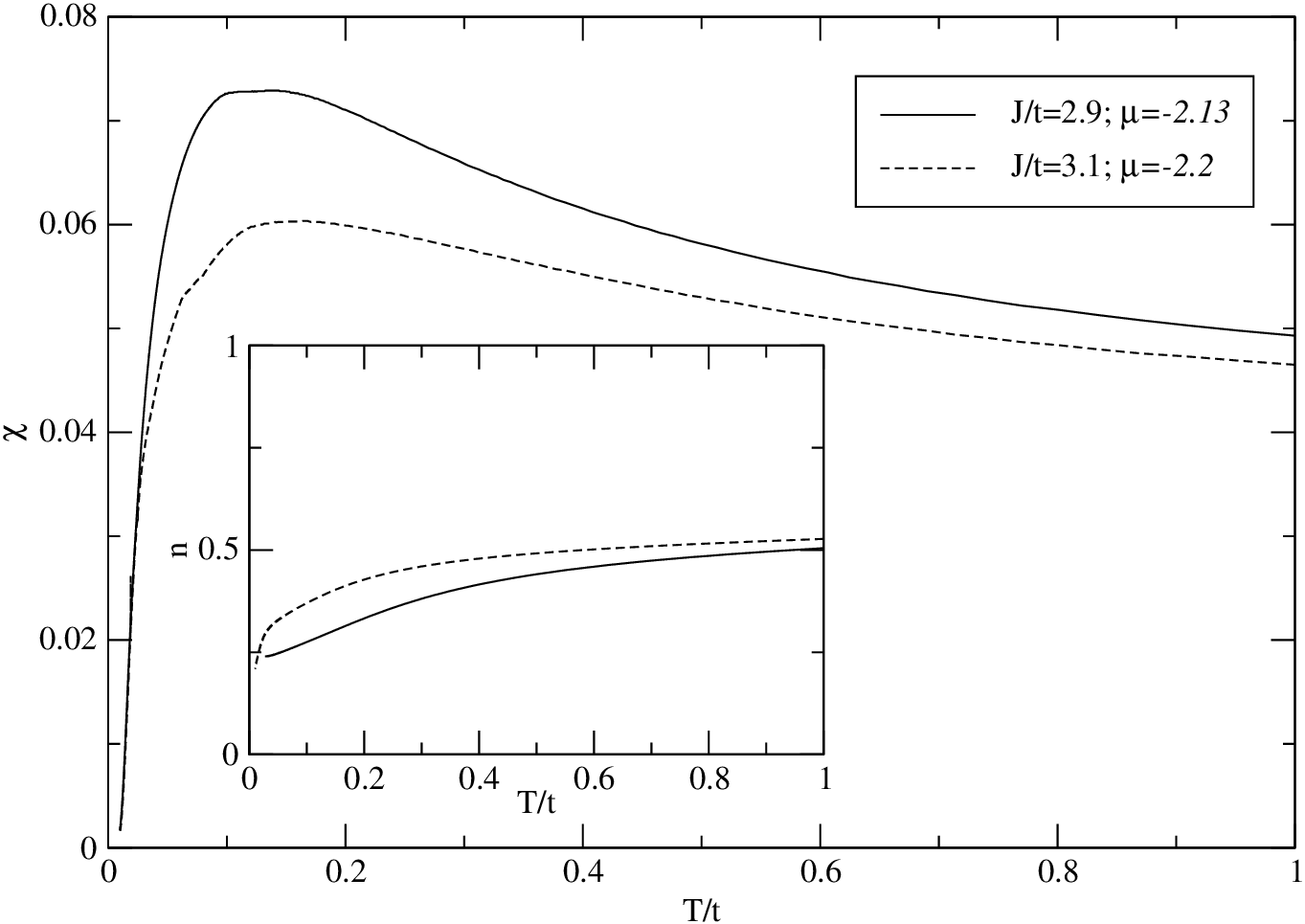}
\caption{Spin susceptibilities for two different parameter sets $(J/t,\mu)$,
  both showing a spin gap of the order $\Delta\sim 0.05$. The inset shows the temperature dependence of the corresponding densities.}
\label{fig8}
\end{figure}
Therefore, we have calculated the spin susceptibility for two different values
of $J/t$ close to phase separation and have chosen chemical potentials so that
the density is given by $n\approx 0.2$ for low temperatures (see
Fig.~\ref{fig8}). In both cases a very small spin gap appears. The quadratic
dispersion of a gapped 1D system leads to $\chi\sim \exp(-\Delta/T)/\sqrt T$
for the low-temperature asymptotics. Using this function for a fit of the numerical data, we find $\Delta=0.05\pm 0.01$ in both cases. 
\begin{figure}[!ht]
\includegraphics*[width=0.9\columnwidth]{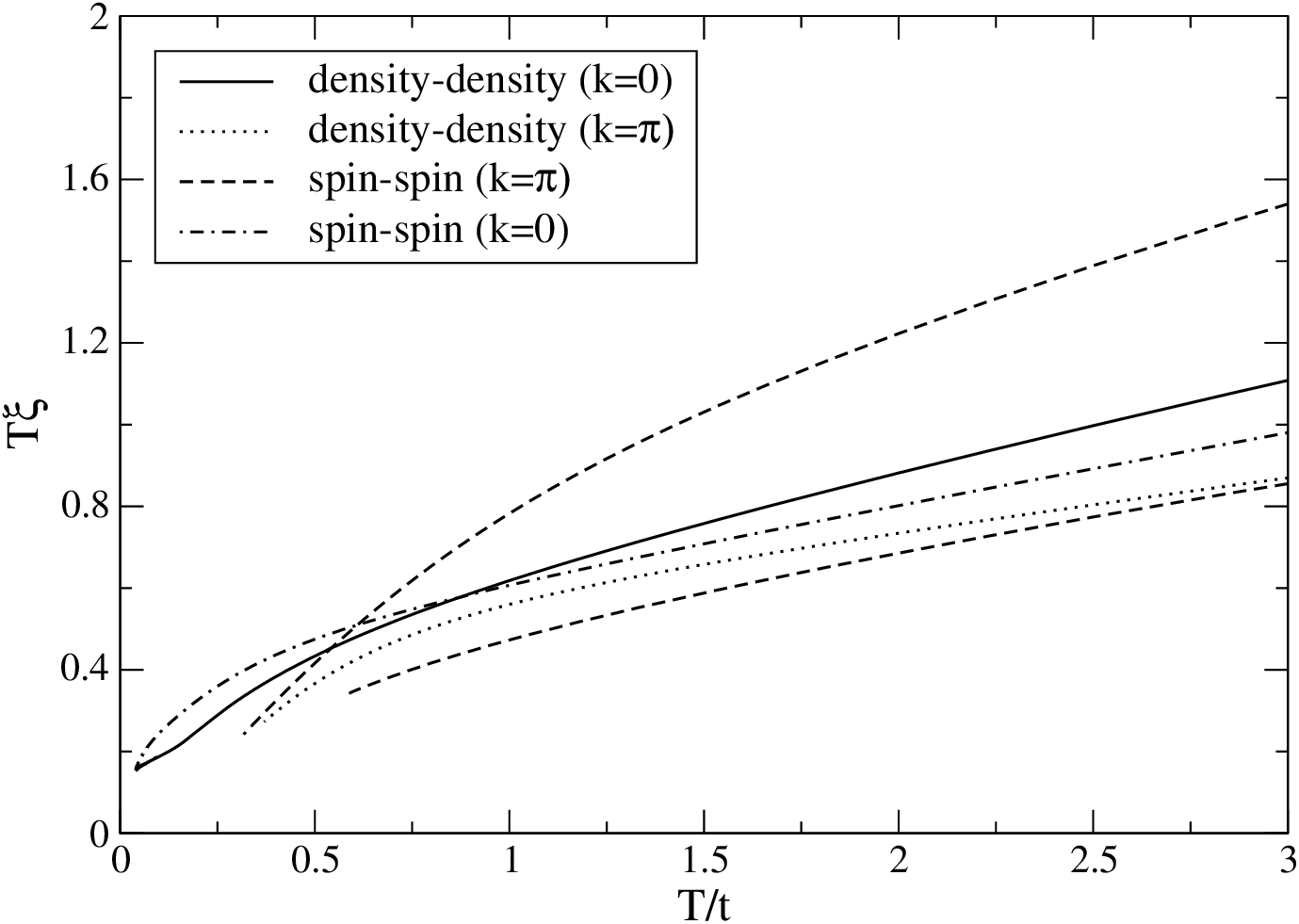}
\caption{Correlation lengths at $J/t=2.9$ and $\mu=-2.13$. Note, that the oscillations of all shown correlation lengths are commensurate over the entire temperature range.}
\label{fig8.1}
\end{figure}

Another proof of LE-properties of the t-J model is given by the calculation of s-s and d-d CL's. In Fig.~\ref{fig8.1} these CL's are shown as usual as temperature times CL versus temperature. A spin gap is connected with finite s-s CL's $\xi_s$ and we therefore expect that $T\cdot \xi_s \rightarrow 0$ for $T\rightarrow 0$. On the other hand the d-d CL's $\xi_n$ are not affected and should still diverge as $\xi_n\sim 1/T$ for low temperatures. This picture seems to be consistent with the numerical results. However, we are not able to present numerical data for lower temperatures, which could support this scenario further.  
\section{t-J Model with Ising anisotropy}
\label{Aniso}
In this section we want to consider the t-J model with modification of Hamiltonian (\ref{t-J_Ham}) by replacing
\begin{equation}
\label{aniso_Ham}
\vec{S}_i\vec{S}_{i+1} \rightarrow S^x_i S^x_{i+1} + S^y_i S^y_{i+1} + \Delta S^z_i S^z_{i+1} - \frac{\Delta}{4} n_i n_{i+1}
\end{equation}
where $\Delta > 1$. We expect that such an anisotropy enhances superconducting
correlations relative to d-d correlations as has been explicitly shown in an exactly
solvable anisotropic t-J model by Bariev {\it et al.}~\cite{BarievKlumper}
However, their version includes some unphysical parity breaking terms (keeping
$PT$-symmetry) making it interesting to investigate if the same is true for
the model defined here. For the Heisenberg chain it is known that an
Ising-like anisotropy promotes long range spin order and causes a spin
gap. The situation is more complicated in the t-J model: The charge sector is
unaffected and the charge excitations remain critical (i.e.~d-d correlations
decay algebraically in the ground state). The expected long range spin order
could be hidden when the density $n \neq 1$, because the spin operators act on
the physical lattice whereas the spins are coupled to the ``electron
lattice''. The long range order would then be visible only in a string order
parameter $\langle \tilde{S}^z_0 \tilde{S}^z_r \rangle$ where all empty sites
between $0$ and $r$ are omitted. This has been emphasized also by Pruschke and
Shiba \cite{PruschkeShiba} who have studied the limit $J/t\rightarrow 0$. In any case there has to be a spin gap also
in the t-J model if $\Delta > 1$. However, in the Heisenberg chain the gap is
given by $\Delta E \propto \exp\left\{-J/(\Delta-1)\right\}$ leading to an
exponential small gap and this gap is further reduced in the t-J model for
densities $n < 1$ making it often undetectable in the thermodynamical data.
\begin{figure}[!ht]
\includegraphics*[width=0.9\columnwidth]{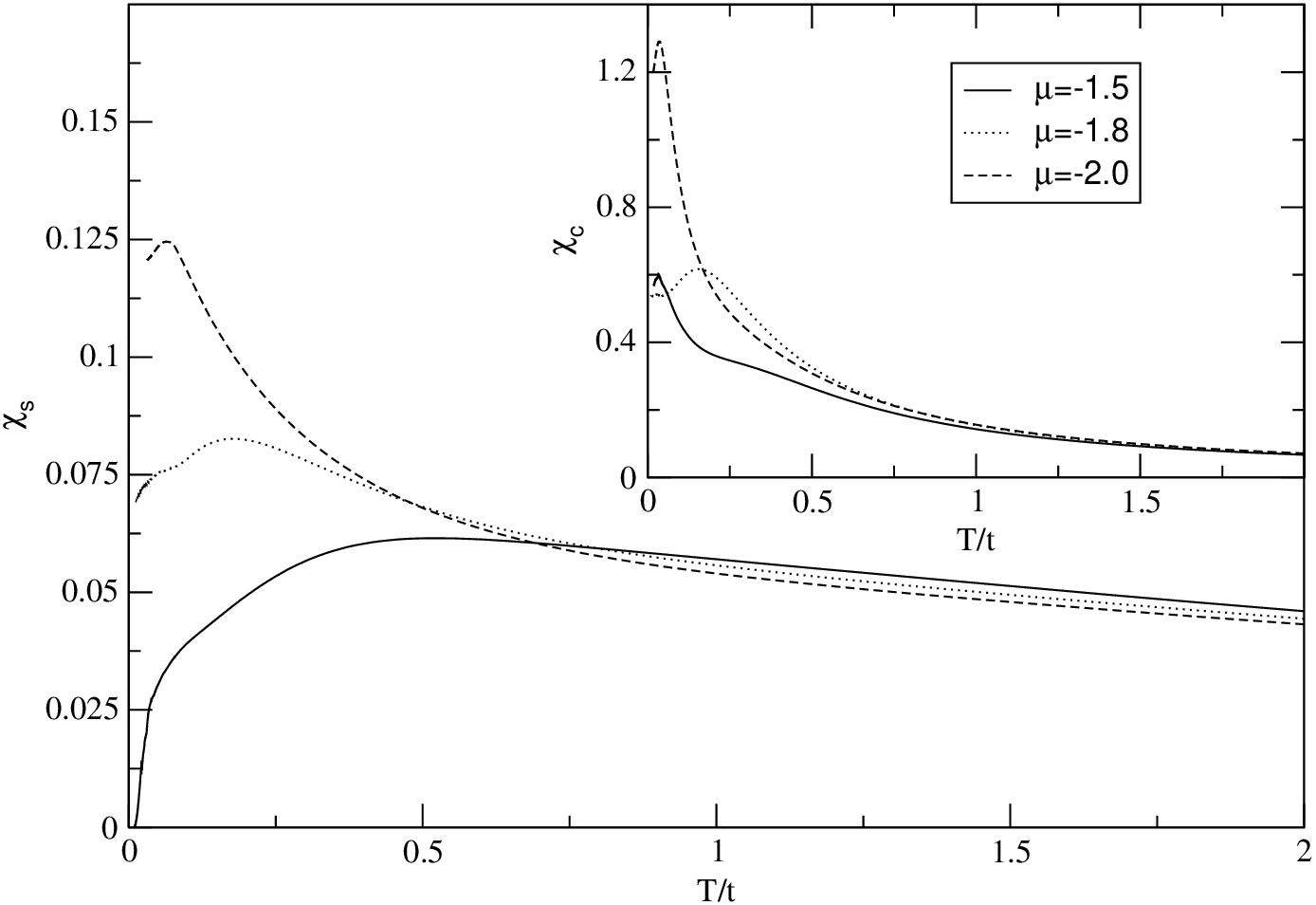}
\caption{Susceptibilities (main figure) and compressibilities (inset) for three different chemical potentials where $J/t=2.0$ and the anisotropy parameter is given by $\Delta=1.5$.}
\label{Aniso_suscept}
\end{figure}
In Fig.~\ref{Aniso_suscept} a spin gap $\Delta_s=0.054\pm0.01$ is visible in
the susceptibility data for $\mu=-1.5$ corresponding to $n \approx 0.98$ in
the low-temperature limit. This is in agreement with the gap $\Delta^{HB}=
0.043$ for the Heisenberg chain with the same anisotropy. For $\mu=-1.8$
($n_{T\rightarrow 0}\approx 0.54$) and $\mu=-2.0$ ($n_{T\rightarrow 0}\approx 0.27$) the spin gap is so small that it is not visible in the accessible temperature range.
\begin{figure}[!ht]
\includegraphics*[width=0.9\columnwidth]{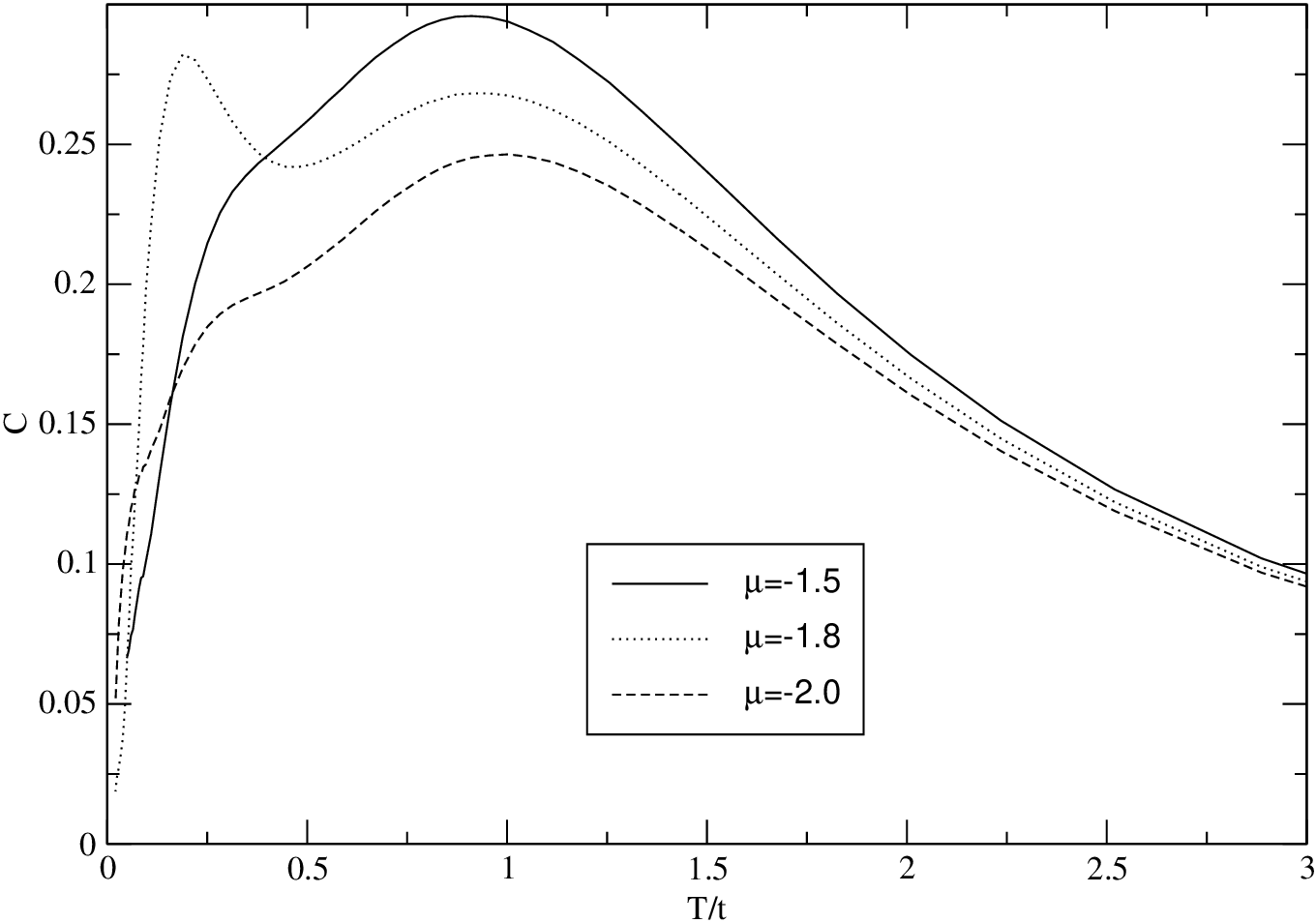}
\caption{Specific heats for the same chemical potentials and same parameters as in Fig.~\ref{Aniso_suscept}.}
\label{Aniso_specHeat}
\end{figure}
The specific heat (Fig.~\ref{Aniso_specHeat}) looks qualitatively similar to the isotropic case. Again two peaks (shoulders) are visible corresponding to spinon and holon excitations.
\begin{figure}[!ht]
\includegraphics*[width=0.9\columnwidth]{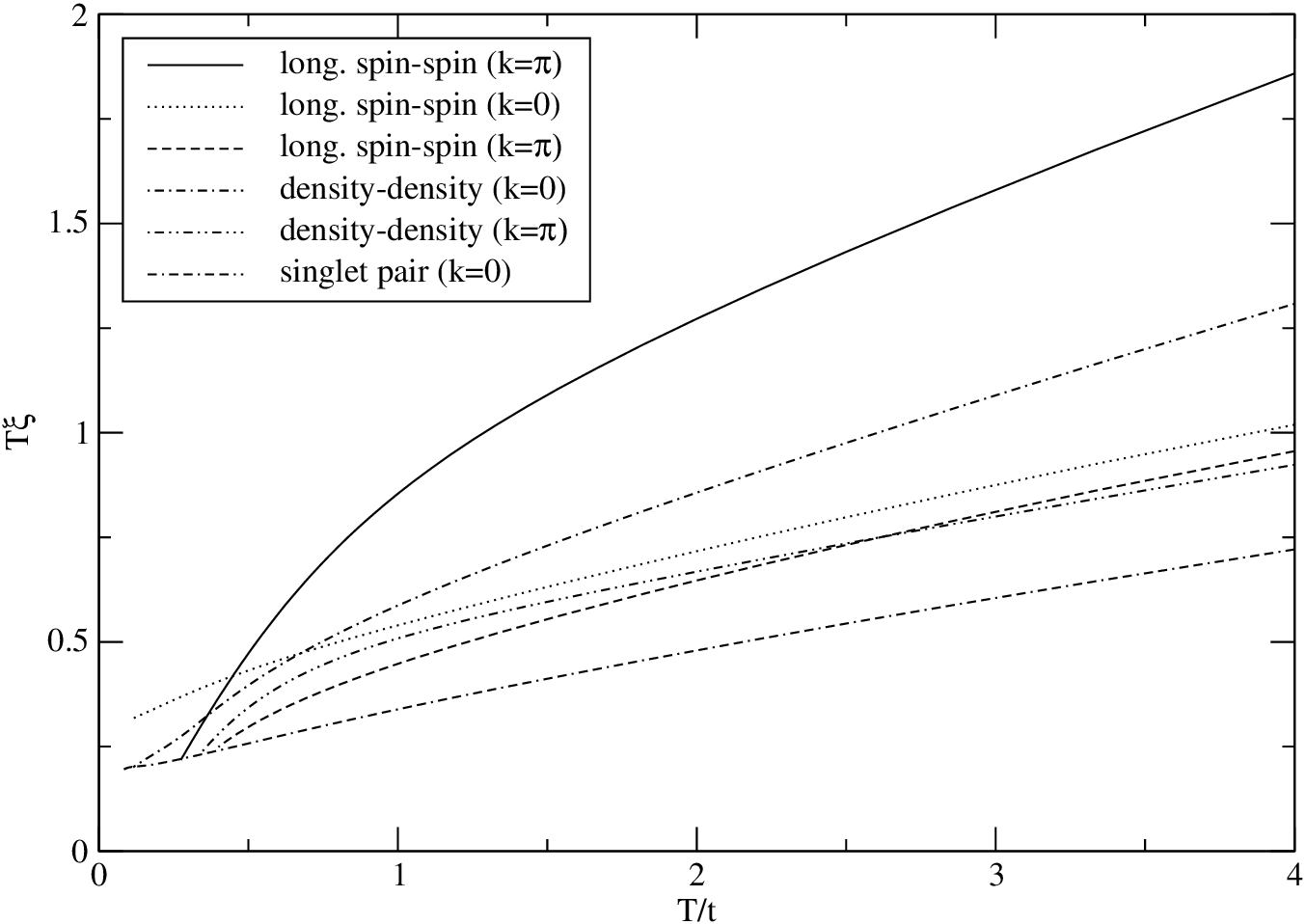}
\caption{Leading d-d and longitudinal s-s CL's for $\mu=-1.8$ and $\Delta=1.5$. Additionally, the leading singlet pair CL is plotted. The shown CL's have commensurate oscillations.}
\label{Aniso_corrlength}
\end{figure}
In Fig.~\ref{Aniso_corrlength} several crossovers in the CL's are visible. In the low-temperature regime the leading d-d and longitudinal s-s CL's diverge as $\xi\sim1/T$ showing that the charge excitations remain critical and that the long-range spin order is hidden. As expected the singlet pair CL dominates at low temperatures over the d-d CL supporting that an Ising-like anisotropy could be a mechanism leading to superconductivity in higher dimensions.\cite{BarievKlumper} 
\section{Conclusions}
\label{Conclusion}
We have described briefly two variants of the TMRG algorithm, where the novel one allows to determine wavevectors unambiguously solely from the eigenvalues of the QTM and makes the calculation of static correlation functions much easier. At the supersymmetric point of the one-dimensional t-J model an excellent agreement between the numerical results and Bethe ansatz was found. Additionally, the numerical results for correlation lengths and static correlation functions have been successfully tested for free spinless fermions. This shows that the TMRG is well suited even for fermionic systems and yields accurate results over a wide temperature range what was sometimes questioned before.\cite{RiceTroyer} In particular, we have concentrated on the calculation of correlation lengths at the supersymmetric point to study the crossover from the non-universal high T lattice into the quantum critical regime ($T\ll t$). The non-universal regime is characterized by various crossovers between CL's with different wavevectors whereas the CL's are non-crossing and diverging as $1/T$ in the universal TLL regime. A good coincidence between predictions for the low-temperature asymptotics of CL's by CFT and the numerical results was shown, but it was important to take also the logarithmic corrections into account.   
For $J/t=0.35$, a value often considered physically relevant, we observed properties that were rather similar to the supersymmetric case. In particular, the t-J model at this parameter point belongs also to the TLL universality class.
In our grand-canonical description phase separation has the meaning of phase coexistence at one chemical potential $\mu$. This is due to the remaining competition between the Heisenberg and the chemical potential term. If $J/t$ is large enough so that total phase separation occurs this competition leads to an empty state for $T=0$ if $\mu<-J\ln 2$ and to a state with $n=1$ if $\mu>-J\ln 2$. Our data also show that the fully phase separated state is destroyed by introducing holes into the Heisenberg island.
By calculating directly the spin susceptibility for small densities and values of $J/t$ near phase separation, we have proven the existence of a spin-gap phase without making additional assumptions. This was also supported by the calculation of s-s and d-d CL's, indicating that all spin CL's are finite at zero temperature, whereas the d-d CL's are unaffected and diverging still as $1/T$. 
Finally we have studied the t-J model with an additional Ising-like anisotropy and have shown that singlet pair correlations are enhanced so that a tendency towards superconductivity is expected in higher dimensions. Further on, we have settled that the expected long range spin order is hidden away from half-filling, because the spins are coupled to the ``electron lattice'' whereas the operators act on the physical lattice.  
\begin{acknowledgments}
J.S.~thanks Ch.~Scheeren for communicating his results from Bethe ansatz and is also grateful for valuable discussions with A.~Kemper. This work is supported by the DFG in SP1073.
\end{acknowledgments}
\appendix
\section{Free spinless fermions}
\label{Appendix}
Here we calculate the density-density correlation function for a system of free spinless fermions on a lattice. The Hamiltonian is given by
\begin{equation}
\label{freeFerm1}
H=-t\sum_q \cos(q) c^\dagger_q c_q
\end{equation}
and the density operator by
\begin{equation}
\label{freeFerm2}
\rho_q(t)=\frac{1}{V}\sum_k c^\dagger_{k+q/2}(t) c_{k-q/2}(t) \: .
\end{equation}
By applying Wick's theorem the d-d correlation function separates into two-point functions
\begin{eqnarray}
\label{freeFerm3}
\!\!\!\!\!\!\!\!&&\left< c_r^\dagger(t)c_r(t)c_0^\dagger(0)c_0(0)\right> = \frac{1}{V^2} \left( \sum_k n(\epsilon_k) \right)^2 \\
&+&\frac{1}{V^2} \sum_{k,q} \e^{\im qr} \left< c^\dagger_{k-\frac{q}{2}}(t) c_{k-\frac{q}{2}}(0) \right> \left< c_{k+\frac{q}{2}}(t) c^\dagger_{k+\frac{q}{2}}(0) \right> \nonumber
\end{eqnarray}
where $\epsilon_k = -t\cos(k)$ and $n(\epsilon_k)$ is the Fermi function. By using the Matsubara formalism, inserting the known results for the electrons Green's function and then going back to real time, this is easily transformed into 
\begin{eqnarray}
\label{freeFerm4}
\!\!\!\!\!\!\!\!&&\left< c_r^\dagger(t)c_r(t)c_0^\dagger(0)c_0(0)\right> = n^2 \\
&+& \frac{1}{V^2}\sum_{q,k} \e^{\im qr} \frac{\e^{\im(\epsilon_{k-\frac{q}{2}}-\epsilon_{k+\frac{q}{2}})t}}{(\e^{\beta\epsilon_{k-\frac{q}{2}}}+1)(\e^{-\beta\epsilon_{k+\frac{q}{2}}}+1)} \: . \nonumber
\end{eqnarray}
Regarding only the connected part and transforming to Matsubara frequencies we get in the free fermion case
\begin{equation}
\label{freeFerm4.1}
G(r,z=0) = \frac{1}{(2\pi)^2}\!\int\!\!\!\int\! dq\, dk\, \frac{n(\epsilon_{k+\frac{q}{2}})-n(\epsilon_{k-\frac{q}{2}})}{\epsilon_{k-\frac{q}{2}}-\epsilon_{k+\frac{q}{2}}}\cos(qr)
\end{equation}
for the static density-density correlation function as defined in Eq.~(\ref{def_static}). 
\begin{figure}[!ht]
\includegraphics*[width=0.9\columnwidth]{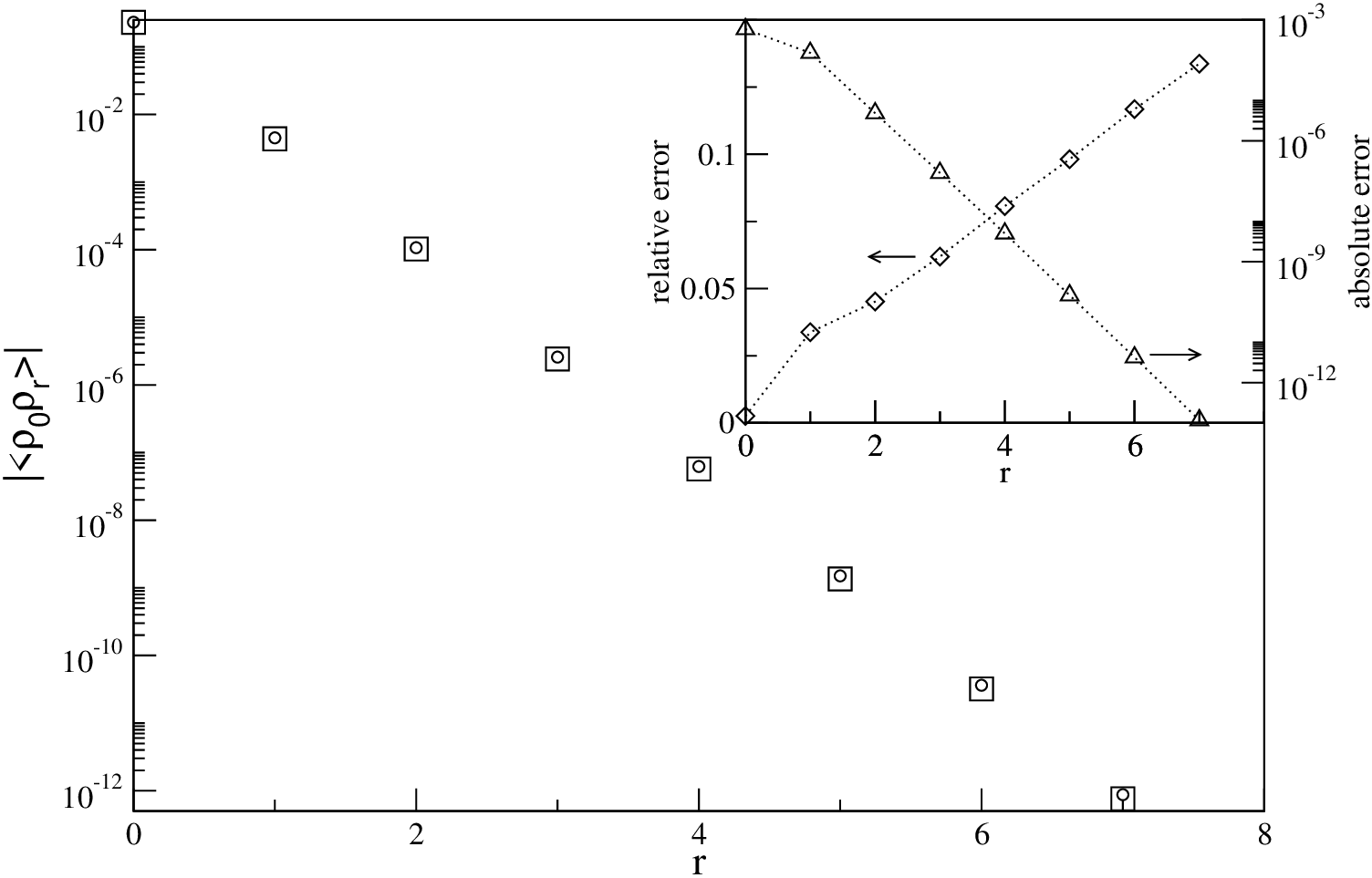}
\includegraphics*[width=0.9\columnwidth]{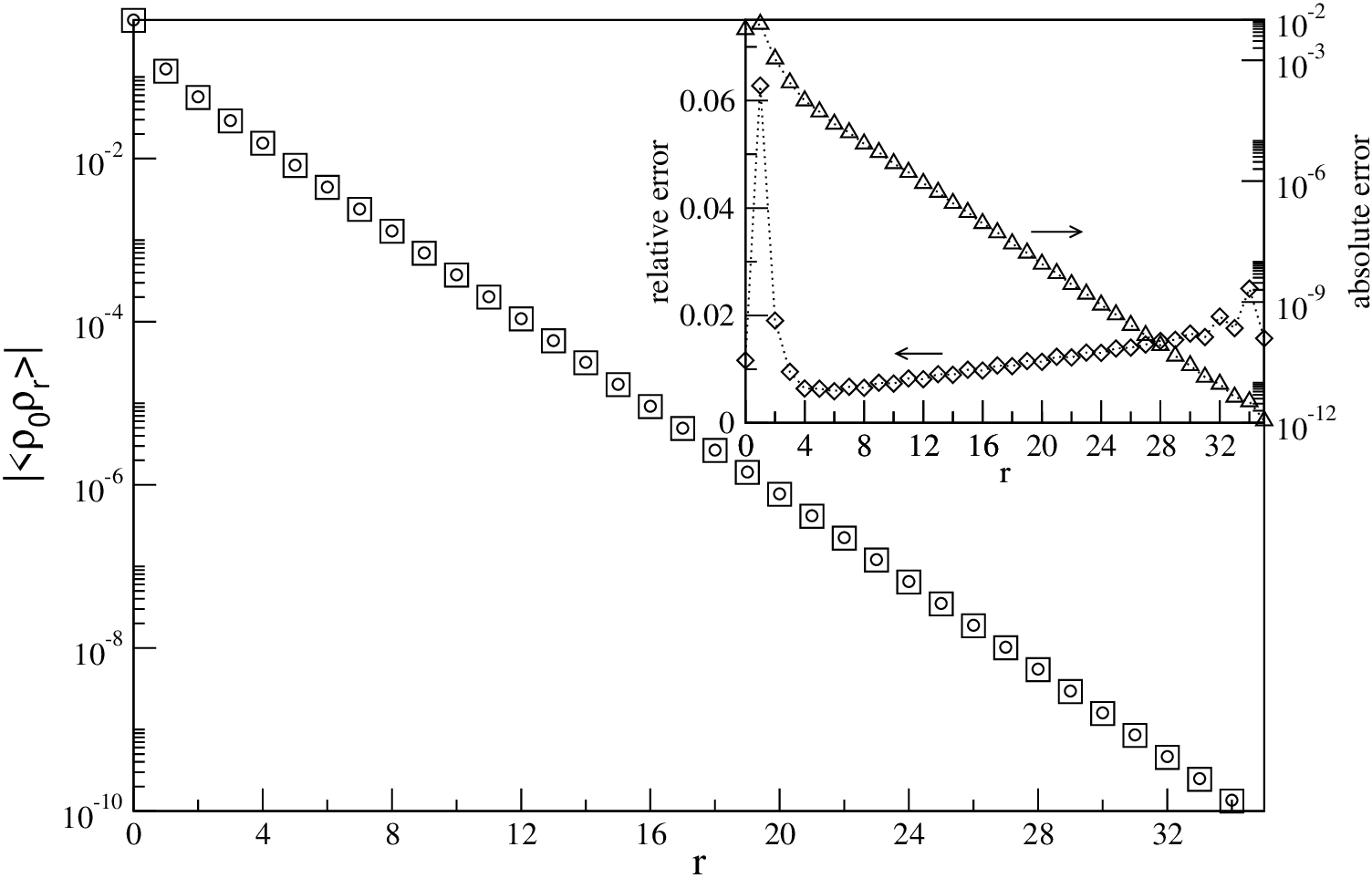}
\caption{The upper (lower) graph shows the static density-density correlation function for free fermions at $T/t=1.0$ ($T/t=0.1$) calculated by Eq.~(\ref{freeFerm4.1}) (circles) and by TMRG (squares) with $N=120$ states retained. The correlation function shows in both cases commensurate oscillations with $k=\pi$. The insets show the absolute errors (triangles) and the relative errors (diamonds) of the TMRG results. The lines are guides to the eye.}
\label{free_fermions.fig1}
\end{figure}
In Fig.~\ref{free_fermions.fig1} we show results for two characteristic temperatures obtained by TMRG using Eqs.~(\ref{static2}) and (\ref{static3}) in comparison to the exact results from a numerical evaluation of Eq.~(\ref{freeFerm4.1}). The absolute error is largest for short distances and decreases exponentially with distance. However, even at $T/t=0.1$ the error always remains smaller than $10^{-2}$ showing that TMRG yields useful results not only for the asymptotic behavior of static correlation functions but also at short distances.  

\begin{figure}[!ht]
\includegraphics*[width=0.9\columnwidth]{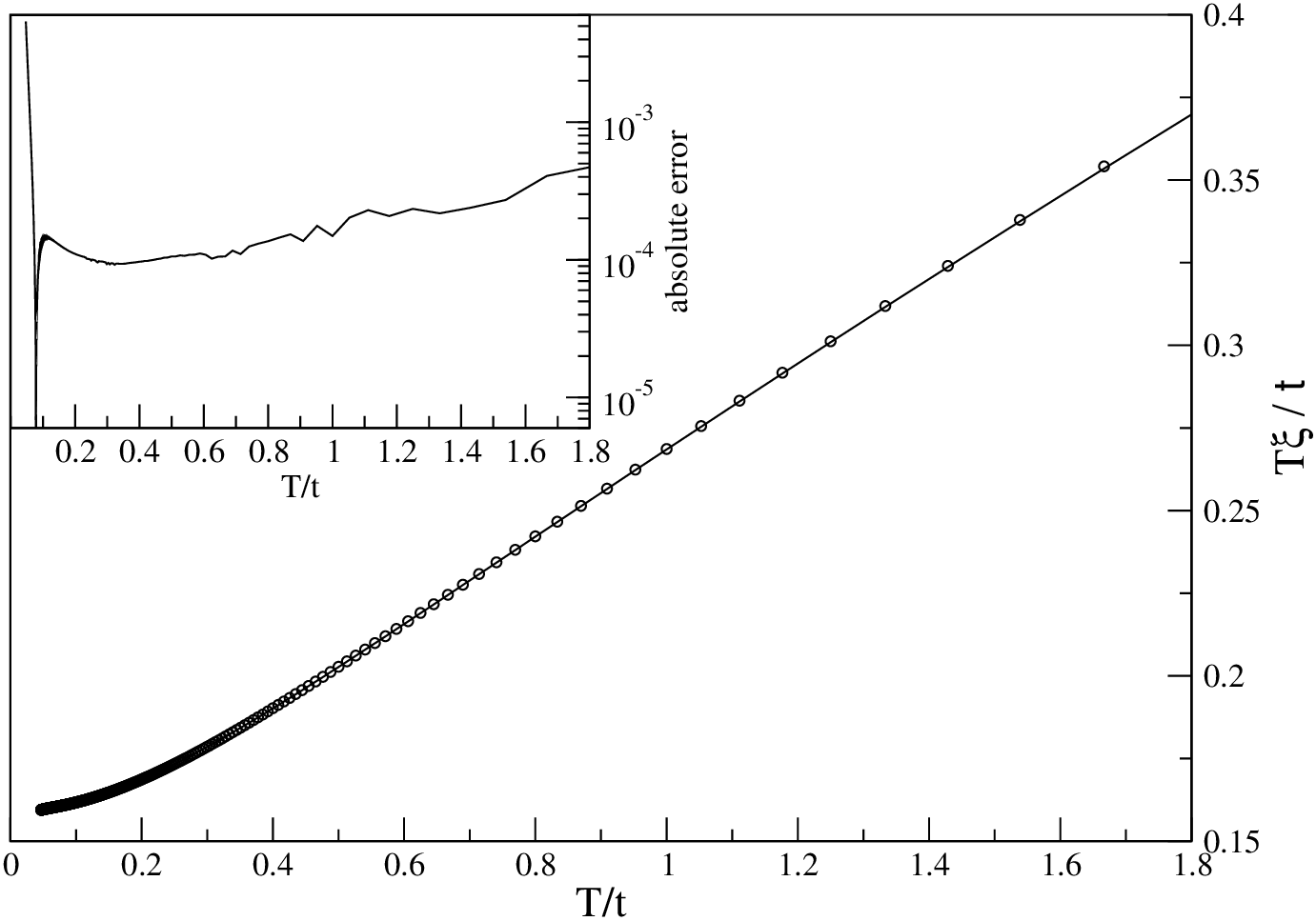}
\caption{Results for the correlation length calculated by Eq.~(\ref{freeFerm9}) (line) compared to TMRG results (circles) with 120 states retained. The inset shows the errors of the TMRG results.}
\label{free_fermions.fig3}
\end{figure}

Returning to Eq.~(\ref{freeFerm4}) and setting $t=0$ we find
\begin{eqnarray}
\label{freeFerm5}
\left< c_r^\dagger c_r c_0^\dagger c_0 \right> &=& \left[\frac{1}{2\pi}\int_{-\pi}^\pi dq \frac{\e^{\im qr}}{\e^{-\beta\cos(q)}+1}\right] \nonumber \\
&\times& \left[\frac{1}{2\pi}\int_{-\pi}^\pi dk \frac{\e^{-\im kr}}{\e^{\beta\cos(k)}+1}\right] \: .
\end{eqnarray}
Now we can in principle evaluate these integrals by closing the integration path in the upper (lower) half plane and using Cauchy's formula. The poles are at the points
\begin{equation}
\label{freeFerm6} 
q,k = \pm \frac{\pi}{2} \pm \im \:\text{arcsinh}\left((2n+1)\frac{\pi}{\beta}\right) 
\end{equation}
and thus, the dominant contribution is 
\begin{subequations}
\begin{equation}
\label{freeFerm7} 
\left< c_r^\dagger c_r c_0^\dagger c_0 \right> \sim M \e^{-r/\xi} (1-\cos(\pi r))
\end{equation}
where the matrixelement is given by
\begin{equation}
\label{freeFerm8}
M = \frac{2}{\pi^2+\beta^2}
\end{equation}
and the CL by
\begin{equation}
\label{freeFerm9}
\xi = \frac{1}{2\:\text{arcsinh}(\pi/\beta)} \: .
\end{equation}
\end{subequations}
In Fig.~\ref{free_fermions.fig3} the leading CL received by Eq.~(\ref{freeFerm9}) is shown together with TMRG results calculated by Eq.~(\ref{corr-length}). A good agreement is obtained with errors of the numerical data remaining smaller than $10^{-3}$ for temperatures down to $T/t\sim 0.1$.  


\begin{thebibliography}{22}
\expandafter\ifx\csname natexlab\endcsname\relax\def\natexlab#1{#1}\fi
\expandafter\ifx\csname bibnamefont\endcsname\relax
  \def\bibnamefont#1{#1}\fi
\expandafter\ifx\csname bibfnamefont\endcsname\relax
  \def\bibfnamefont#1{#1}\fi
\expandafter\ifx\csname citenamefont\endcsname\relax
  \def\citenamefont#1{#1}\fi
\expandafter\ifx\csname url\endcsname\relax
  \def\url#1{\texttt{#1}}\fi
\expandafter\ifx\csname urlprefix\endcsname\relax\def\urlprefix{URL }\fi
\providecommand{\bibinfo}[2]{#2}
\providecommand{\eprint}[2][]{\url{#2}}

\bibitem[{\citenamefont{Ogata et~al.}(1991)\citenamefont{Ogata, Luchini,
  Sorella, and Assaad}}]{OgataLuchini}
\bibinfo{author}{\bibfnamefont{M.}~\bibnamefont{Ogata}},
  \bibinfo{author}{\bibfnamefont{M. U.}~\bibnamefont{Luchini}},
  \bibinfo{author}{\bibfnamefont{S.}~\bibnamefont{Sorella}}, \bibnamefont{and}
  \bibinfo{author}{\bibfnamefont{F. F.}~\bibnamefont{Assaad}},
  \bibinfo{journal}{Phys. Rev. Lett.} \textbf{\bibinfo{volume}{66}},
  \bibinfo{pages}{2388} (\bibinfo{year}{1991}).

\bibitem[{\citenamefont{Hellberg and Mele}(1991)}]{HellbergMele}
\bibinfo{author}{\bibfnamefont{C. S.}~\bibnamefont{Hellberg}} \bibnamefont{and}
  \bibinfo{author}{\bibfnamefont{E. J.}~\bibnamefont{Mele}},
  \bibinfo{journal}{Phys. Rev. Lett.} \textbf{\bibinfo{volume}{67}},
  \bibinfo{pages}{2080} (\bibinfo{year}{1991}).

\bibitem[{\citenamefont{Bares and Blatter}(1990)}]{BaresBlatter}
\bibinfo{author}{\bibfnamefont{P. A.}~\bibnamefont{Bares}} \bibnamefont{and}
  \bibinfo{author}{\bibfnamefont{G.}~\bibnamefont{Blatter}},
  \bibinfo{journal}{Phys. Rev. Lett.} \textbf{\bibinfo{volume}{64}},
  \bibinfo{pages}{2567} (\bibinfo{year}{1990}).

\bibitem[{\citenamefont{Kawakami and Yang}(1990)}]{KawakamiYang}
\bibinfo{author}{\bibfnamefont{N.}~\bibnamefont{Kawakami}} \bibnamefont{and}
  \bibinfo{author}{\bibfnamefont{S. K.}~\bibnamefont{Yang}},
  \bibinfo{journal}{Phys. Rev. Lett.} \textbf{\bibinfo{volume}{65}},
  \bibinfo{pages}{2309} (\bibinfo{year}{1990}),
\bibinfo{author}{\bibfnamefont{N.}~\bibnamefont{Kawakami}} \bibnamefont{and}
  \bibinfo{author}{\bibfnamefont{S. K.}~\bibnamefont{Yang}}, \bibinfo{journal}{J.
  Phys. Cond. Mat.} \textbf{\bibinfo{volume}{3}}, \bibinfo{pages}{5983}
  (\bibinfo{year}{1991}).

\bibitem[{\citenamefont{J\"uttner et~al.}(1996)\citenamefont{J\"uttner,
  Kl\"umper, and Suzuki}}]{JuettnerKluemper}
\bibinfo{author}{\bibfnamefont{G.}~\bibnamefont{J\"uttner}},
  \bibinfo{author}{\bibfnamefont{A.}~\bibnamefont{Kl\"umper}},
  \bibnamefont{and} \bibinfo{author}{\bibfnamefont{J.}~\bibnamefont{Suzuki}},
  \bibinfo{journal}{Nuc. Phys. B}  \textbf{\bibinfo{volume}{487}},
\bibinfo{pages}{650} (\bibinfo{year}{1997}).

\bibitem[{\citenamefont{Chen and Lee}(1996)}]{ChenLee}
\bibinfo{author}{\bibfnamefont{Y. C.}~\bibnamefont{Chen}} \bibnamefont{and}
  \bibinfo{author}{\bibfnamefont{T. K.}~\bibnamefont{Lee}},
  \bibinfo{journal}{Phys. Rev. B} \textbf{\bibinfo{volume}{54}},
  \bibinfo{pages}{9062} (\bibinfo{year}{1996}).

\bibitem[{\citenamefont{Kobayashi et~al.}(1996)\citenamefont{Kobayashi, Ohe,
  and Iguchi}}]{KobayashiOhe}
\bibinfo{author}{\bibfnamefont{K.}~\bibnamefont{Kobayashi}},
  \bibinfo{author}{\bibfnamefont{C.}~\bibnamefont{Ohe}}, \bibnamefont{and}
  \bibinfo{author}{\bibfnamefont{K.}~\bibnamefont{Iguchi}},
  \bibinfo{journal}{Phys. Rev. B} \textbf{\bibinfo{volume}{54}},
  \bibinfo{pages}{13129} (\bibinfo{year}{1996}).

\bibitem[{\citenamefont{Nakamura et~al.}(1997)\citenamefont{Nakamura, Nomura,
  and Kitazawa}}]{NakamuraNomura}
\bibinfo{author}{\bibfnamefont{M.}~\bibnamefont{Nakamura}},
  \bibinfo{author}{\bibfnamefont{K.}~\bibnamefont{Nomura}}, \bibnamefont{and}
  \bibinfo{author}{\bibfnamefont{A.}~\bibnamefont{Kitazawa}},
  \bibinfo{journal}{Phys. Rev. Lett.} \textbf{\bibinfo{volume}{79}},
  \bibinfo{pages}{3214} (\bibinfo{year}{1997}).

\bibitem[{\citenamefont{Okiji and Suga}(1997)}]{OkijiSuga}
\bibinfo{author}{\bibfnamefont{A.}~\bibnamefont{Okiji}} \bibnamefont{and}
  \bibinfo{author}{\bibfnamefont{S.}~\bibnamefont{Suga}},
  \bibinfo{journal}{Physica B} \textbf{\bibinfo{volume}{237}},
  \bibinfo{pages}{81} (\bibinfo{year}{1997}).

\bibitem[{\citenamefont{Frahm and Korepin}(1990)}]{FrahmKorepin}
\bibinfo{author}{\bibfnamefont{H.}~\bibnamefont{Frahm}} \bibnamefont{and}
  \bibinfo{author}{\bibfnamefont{V. E.}~\bibnamefont{Korepin}},
  \bibinfo{journal}{Phys. Rev. B} \textbf{\bibinfo{volume}{42}},
  \bibinfo{pages}{10553} (\bibinfo{year}{1990}).

\bibitem[{\citenamefont{Bursill et~al.}(1996)\citenamefont{Bursill, Xiang, and
  Gehring}}]{BursillXiang}
\bibinfo{author}{\bibfnamefont{R.}~\bibnamefont{Bursill}},
  \bibinfo{author}{\bibfnamefont{T.}~\bibnamefont{Xiang}}, \bibnamefont{and}
  \bibinfo{author}{\bibfnamefont{G.}~\bibnamefont{Gehring}},
  \bibinfo{journal}{J. Phys. Cond. Mat.} \textbf{\bibinfo{volume}{8}},
  \bibinfo{pages}{L583} (\bibinfo{year}{1996}).

\bibitem[{\citenamefont{Wang and Xiang}(1997)}]{WangXiang}
\bibinfo{author}{\bibfnamefont{X.}~\bibnamefont{Wang}} \bibnamefont{and}
  \bibinfo{author}{\bibfnamefont{T.}~\bibnamefont{Xiang}},
  \bibinfo{journal}{Phys. Rev. B} \textbf{\bibinfo{volume}{56}},
  \bibinfo{pages}{5061} (\bibinfo{year}{1997}).

\bibitem[{\citenamefont{Shibata}(1997)}]{Shibata}
\bibinfo{author}{\bibfnamefont{N.}~\bibnamefont{Shibata}}, \bibinfo{journal}{J.
  Phys. Soc. Jap.} \textbf{\bibinfo{volume}{66}}, \bibinfo{pages}{2221}
  (\bibinfo{year}{1997}).

\bibitem[{\citenamefont{Kl\"umper et~al.}(1999)\citenamefont{Kl\"umper,
  Raupach, and Sch\"onfeld}}]{Raupach}
\bibinfo{author}{\bibfnamefont{A.}~\bibnamefont{Kl\"umper}},
  \bibinfo{author}{\bibfnamefont{R.}~\bibnamefont{Raupach}}, \bibnamefont{and}
  \bibinfo{author}{\bibfnamefont{F.}~\bibnamefont{Sch\"onfeld}},
  \bibinfo{journal}{Phys. Rev. B} \textbf{\bibinfo{volume}{59}},
  \bibinfo{pages}{3612} (\bibinfo{year}{1999}).

\bibitem[{\citenamefont{Sirker and Kl\"umper}(2002)}]{SirkerKluemper2}
\bibinfo{author}{\bibfnamefont{J.}~\bibnamefont{Sirker}} \bibnamefont{and}
  \bibinfo{author}{\bibfnamefont{A.}~\bibnamefont{Kl\"umper}},
  \bibinfo{journal}{Europhys. Lett.} \textbf{\bibinfo{volume}{60}},
  \bibinfo{pages}{262} (\bibinfo{year}{2002}).

\bibitem[{\citenamefont{Giamarchi and Schulz}(1989)}]{GiamarchiSchulz}
\bibinfo{author}{\bibfnamefont{T.}~\bibnamefont{Giamarchi}} \bibnamefont{and}
  \bibinfo{author}{\bibfnamefont{H. J.}~\bibnamefont{Schulz}},
  \bibinfo{journal}{Phys. Rev. B} \textbf{\bibinfo{volume}{39}},
  \bibinfo{pages}{4620} (\bibinfo{year}{1989}).

\bibitem[{\citenamefont{Kl\"umper et~al.}(2001)\citenamefont{Kl\"umper,
  Martinez, Scheeren, and Shiroishi}}]{KluemperScheeren}
\bibinfo{author}{\bibfnamefont{A.}~\bibnamefont{Kl\"umper}},
  \bibinfo{author}{\bibfnamefont{J.~R.} \bibnamefont{Martinez}},
  \bibinfo{author}{\bibfnamefont{C.}~\bibnamefont{Scheeren}}, \bibnamefont{and}
  \bibinfo{author}{\bibfnamefont{M.}~\bibnamefont{Shiroishi}},
  \bibinfo{journal}{J. Stat. Phys.} \textbf{\bibinfo{volume}{102}},
  \bibinfo{pages}{937} (\bibinfo{year}{2001}).

\bibitem[{\citenamefont{Sakai and Kl\"umper}(2001)}]{SakaiKluemper}
\bibinfo{author}{\bibfnamefont{K.}~\bibnamefont{Sakai}} \bibnamefont{and}
  \bibinfo{author}{\bibfnamefont{A.}~\bibnamefont{Kl\"umper}},
  \bibinfo{journal}{J. Phys. A} \textbf{\bibinfo{volume}{34}},
  \bibinfo{pages}{8015} (\bibinfo{year}{2001}).

\bibitem[{\citenamefont{I.Peschel et~al.}(1999)}]{Peschel}
\bibinfo{editor}{\bibnamefont{I.Peschel}} \bibnamefont{et~al.}, eds.,
  \emph{\bibinfo{title}{Density-Matrix Renormalization}}, \bibinfo{title}{Lecture Notes in Physics}, vol. \bibinfo{volume}{528}
  (\bibinfo{publisher}{Springer, Berlin}, \bibinfo{year}{1999}),
  \bibinfo{note}{and references therein}.

\bibitem[{\citenamefont{Ammon et~al.}(1999)\citenamefont{Ammon, Troyer, Rice,
  and Shibata}}]{RiceTroyer}
\bibinfo{author}{\bibfnamefont{B.}~\bibnamefont{Ammon}},
  \bibinfo{author}{\bibfnamefont{M.}~\bibnamefont{Troyer}},
  \bibinfo{author}{\bibfnamefont{T.~M.} \bibnamefont{Rice}}, \bibnamefont{and}
  \bibinfo{author}{\bibfnamefont{N.}~\bibnamefont{Shibata}},
  \bibinfo{journal}{Phys. Rev. Lett.} \textbf{\bibinfo{volume}{82}},
  \bibinfo{pages}{3855} (\bibinfo{year}{1999}).

\bibitem[{\citenamefont{Pruschke and Shiba}(1991)}]{PruschkeShiba}
\bibinfo{author}{\bibfnamefont{T.}~\bibnamefont{Pruschke}} \bibnamefont{and}
  \bibinfo{author}{\bibfnamefont{H.}~\bibnamefont{Shiba}},
  \bibinfo{journal}{Phys. Rev. B} \textbf{\bibinfo{volume}{44}},
  \bibinfo{pages}{205} (\bibinfo{year}{1991}).

\bibitem[{\citenamefont{Bariev et~al.}(1995)\citenamefont{Bariev, Kl\"umper,
  Schadschneider, and Zittartz}}]{BarievKlumper}
\bibinfo{author}{\bibfnamefont{R.}~\bibnamefont{Bariev}},
  \bibinfo{author}{\bibfnamefont{A.}~\bibnamefont{Kl\"umper}},
  \bibinfo{author}{\bibfnamefont{A.}~\bibnamefont{Schadschneider}},
  \bibnamefont{and} \bibinfo{author}{\bibfnamefont{J.}~\bibnamefont{Zittartz}},
  \bibinfo{journal}{Z. Phys. B} \textbf{\bibinfo{volume}{96}},
  \bibinfo{pages}{395} (\bibinfo{year}{1995}).

\end{thebibliography}
\end{document}